\documentclass[11pt]{article}
\usepackage{amssymb,amsmath}
\numberwithin{equation}{section}
\usepackage{color}
\usepackage{cite}
\usepackage{hyperref}
\usepackage{graphicx}
\usepackage{MnSymbol}
\usepackage[font=footnotesize,labelfont=bf]{caption}
\usepackage[margin=1in]{geometry}
\usepackage{tikz}
\usepackage{mathtools}
\usepackage{tikz-3dplot}
\usetikzlibrary{decorations.markings}

\newcommand{\Hphys}{\mathcal{H}_{\mathrm{phys}}} 

\newcommand{\ibraket}[2]{\lsem #1 | #2 \rsem}

\newcommand{\hinv}{\hilb_\mathrm{inv}} 
\newcommand{\hco}{\hilb_\mathrm{co}} 

\newcommand{\ham}{H} 

\newcommand{\hilb}{\mathcal{H}}	
\newcommand{\ZZ}{\mathbb{Z}}		
\newcommand{\RR}{\mathbb{R}}		
\newcommand{\fin}{\phi_\mathrm{in}}  
\newcommand{\fout}{\phi_\mathrm{out}} 
\newcommand{\finout}{\phi_\mathrm{in/out}}

\newcommand{\li}{\lsem}
\newcommand{\ri}{\rsem}
\newcommand{\lc}{\llangle}
\newcommand{\rc}{\rrangle}

\DeclareMathOperator{\sgn}{sgn}
\DeclareMathOperator{\sech}{sech}

\newcommand{\scri}{\mathcal{I}}

\DeclareMathOperator{\Tr}{Tr}

\title{The Hilbert space of de Sitter JT: \\ a case study for canonical methods in quantum gravity}

\author{Jesse Held\thanks{\href{mailto:jheld@umail.ucsb.edu}{\texttt{jheld@umail.ucsb.edu}}} \\Department of Physics, UC Santa Barbara  \and Henry Maxfield\thanks{\href{mailto:henrym@stanford.edu}{\texttt{henrym@stanford.edu}}} \\Stanford Institute for Thoeretical Physics}

\begin{document}
\maketitle
\begin{abstract}
    We study de Sitter JT gravity in the canonical formulation to illustrate  constructions of Hilbert spaces in quantum gravity, which is challenging due to the Hamiltonian constraints. The key ideas include representing states as `invariants' (solutions to the Wheeler-DeWitt equation) or dual `co-invariants' (equivalence classes under gauge transformations), defining a physical inner product by group averaging, and relating this to Klein-Gordon inner products via gauge-fixing conditions. We identify a rich Hilbert space with positive-definite inner product which splits into distinct sectors, mirroring a similar structure in the classical phase space. Many (but not all) of these sectors are described exactly (in a constant extrinsic curvature gauge) by a mini-superspace theory, a quantum mechanical theory with a single constraint.
\end{abstract}

\tableofcontents


\section{Introduction}

In recent years, path integral methods have proven to be a powerful tool for understanding quantum gravity. However, this formulation is not without its limitations. Notably, gravitational path integrals are simplest to formulate and interpret in situations with asymptotic spatial boundaries. Most prominent among these are asymptotically AdS spacetimes, where we have the additional guidance of the AdS/CFT correspondence to provide an interpretation in terms of a dual non-gravitational system. But this has made it harder to address questions regarding cosmological spacetimes, where there is no asymptotic spatial boundary, or inherently `bulk' questions (such as physics in the black hole interior) which cannot be straightforwardly cast in terms of asymptotic boundary conditions.

This motivates us to revisit the canonical formulation of gravity, in a form which makes local 
bulk physics as manifest as possible, and casts us free from reliance on asymptotic spatial boundaries. Specifically, we would like a Hilbert space of states described by wavefunctions $\psi(\gamma,\Phi)$ depending on bulk quantities (spatial metric $\gamma$ and matter fields $\Phi$), with observables computed by inner products between appropriate states (or matrix elements of operators).\footnote{Our aim is not a complete theory of quantum gravity, but rather an effective description valid at small $\hbar G_N$ (compared to typical scales of the problem). The ideal target would be an expansion in $\hbar G_N$ which at zeroth order is as close as possible to a conventional description of QFT in a curved background spacetime (including free gravitons), with systematic order-by-order perturbation theory, as well as the possible inclusion non-perturbative effects. The latter are our ultimate primary motivation, particularly driven by recent success of spacetime wormholes for capturing details of quantum black holes. We comment further in section \ref{sec:disc}.} In a theory of gravity, the main challenge to such a formulation comes from diffeomorphism invariance (or the associated momentum and Hamiltonian constraints $\mathcal{H}=\mathcal{P}=0$). Apart from the technical complications of gravitational constraints, we must also contend with the conceptual challenge that diffeomorphism invariance can obscure the desired local spacetime description (one facet of the `problem of time'). Perhaps the trickiest and least  known aspect of all is the definition of a sensible inner product taking the constraints into account.

With these motivations in mind, in this paper we study a simple model of cosmology, mostly as an instructive example to illustrate some underappreciated ideas in canonical quantum gravity. This model is the de Sitter version of Jackiw-Teitelboim (JT) gravity \cite{Jackiw:1984je,Teitelboim:1983ux}, a two-dimensional theory of metric $g$ and scalar dilaton field $\Phi$ with action
\begin{equation}
     I = \frac{1}{4\pi}\int d^2x \sqrt{-g} \Phi (\mathcal{R}-2).
     \label{dS Action}
\end{equation}
This theory has also been recently studied in \cite{Maldacena:2019cbz,Cotler:2019dcj,Cotler_2020,Cotler:2023eza,Cotler:2024xzz,Nanda:2023wne}. Our aim is a careful \emph{bulk} description of the Hilbert space (including the inner product), in contrast to much of the previous work which mostly calculated quantities from path integrals defined with asymptotic boundary conditions at timelike future/past infinity $\scri_\pm$. This requires some tools which are not well-known to address some of the conceptual and technical issues outlined above. The simplicity of JT gravity allows us to implement this exactly, so is well-suited to give a very concrete and explicit demonstration of the ideas we wish to explain. We now give a brief outline of the main concepts, which will be explained in more detail in the text. We also explore these topics (and more, such as the detailed relation to BRST quantisation) in yet more detail in parallel work \cite{Held:2025mai}.

While this work was in preparation the paper \cite{Alonso-Monsalve:2024oii} appeared, which has significant overlap with our discussion of the  classical phase space.

\subsection{Summary of main ideas}

\subsubsection{Invariant and co-invariant states}

Perhaps the best-known method for defining physical states is to demand that wavefunctions are annihilated by the constraints, or (for gravity)  the Wheeler-DeWitt equation $\mathcal{H}|\psi\rangle=0$. We call these \emph{invariants} as they are left invariant by gauge transformations. But this is not the only possible approach. An alternative does not impose conditions on the wavefunctions, but instead declares equivalences like $|\psi\rangle\sim |\psi\rangle + \mathcal{H}|\phi\rangle$. We call the resulting equivalence classes \emph{co-invariants} as they are dual to the invariants in the sense of linear algebra.\footnote{This terminology follows \cite{Chandrasekaran_2023}.} Representing states as co-invariants can be useful, particularly for making manifest physics which is local in time. But invariant wavefunctions are also useful, since they arise naturally as states defined by boundary conditions (e.g., at de Sitter timelike infinity $\scri_\pm$, or the Hartle-Hawking state). So it is beneficial to understand both, and the relation between them.\footnote{These are two possible extremes of a much more general set of possibilities, which are unified in the BRST formalism, described in \cite{Held:2025mai} (see also \cite{Chandrasekaran_2023}, \cite{witten2023notecanonicalformalismgravity}).}

\subsubsection{Group averaging} To define an inner product on the physical Hilbert space, we advocate for the \emph{group-averaging} proposal (which falls under the more general framework of `refined algebraic quantisation')\cite{Ashtekar:1995zh,Giulini:1998rk,Giulini_2000,marolf2000groupaveragingrefinedalgebraic}. This aims to define an inner product of co-invariant states by an integral over the gauge group (diffeomorphisms in the case of gravity), which has a natural implementation as a Lorentzian gravitational path integral. This procedure leads to an inner product which is physically sensible, in particular being positive (semi-)definite. We can equivalently think of this inner product as a `rigging map' $\eta$ from co-invariants to invariants:  by averaging an arbitrary wavefunction $|\psi\rangle$ over all possible time translations, we produce a solution $\eta|\psi\rangle$ to the Wheeler-DeWitt equation. By the duality between invariants and co-invariants, this also defines an inner product on invariant (Wheeler-DeWitt) states, which makes these two possible descriptions of Hilbert space equivalent.

\subsubsection{Klein-Gordon inner products and gauge-fixing}

The group-averaging inner product on invariant states has the drawback that it is a little indirect, requiring us to invert the rigging map $\eta$. This is really a generalised inverse ($\kappa$ satisfying $\eta\kappa\eta=\eta$) since a co-invariant state is an equivalence class of wavefunctions, and we have the freedom to choose any representative. It is therefore useful to try to find an explicit construction of such an inverse map $\kappa$.

Such a map can be thought of as arising from a gauge choice, selecting a $t=0$ slice to fix the Hamiltonian constraints (or at least the residual overall time translation which remains from an incomplete gauge-fixing). But it is tricky (particularly for gravity) to pick a local gauge condition that works well for all possible configurations (a `Gribov problem'). For a given spacetime there may be no Cauchy surface satisfying the chosen gauge, or there may be several, or the gauge-invariant measure (from the na\"ive Fadeev-Popov determinant for example) can be negative because the trajectory crosses the gauge slice in the wrong direction. As a result, there is no simple universal choice which reproduces the group-averaging results. Nonetheless, we describe several choices which are useful  for calculating different quantities, and agree with group-averaging in a predictable set of circumstances (at least in perturbation theory).

The well-known `Klein-Gordon' inner product proposed in DeWitt's seminal paper \cite{DeWitt:1967yk} can be though of as  an example of this gauge-fixing, where the gauge choice corresponds to a hypersurface in `superspace'.\footnote{Here we mean Wheeler's superspace of all possible spatial geometries and field configurations, unrelated to the  supersymmetric construction.} It is well-known that this does not give a satisfactory inner product, particularly because it fails to be positive-definite. We explain this as arising from the failure to be a universally good gauge --- `negative frequency' solutions with negative norm arise from configurations which cross the gauge-fixing surface in the wrong direction giving the wrong sign to the measure. Additionally, zero norm states can occur from configurations which do not encounter the gauge surface, or cross it twice in opposite directions (such as in universes which expand and recollapse). Nonetheless, on positive frequency solutions (when such a notion makes sense), the Klein-Gordon inner product agrees (at least in perturbation theory) with group-averaging, while in many cases being more straightforward to calculate.

We also describe alternatives to the Klein-Gordon inner product, most notably the gauge choice of fixing to a geodesic Cauchy surface. This is analogous to the `maximal volume gauge' recently reviewed and studied in \cite{witten2023notecanonicalformalismgravity}.

\subsubsection{Non--self-adjoint constraints} Classical gravity often evolves to singularities in finite proper time --- examples are the physical singularities inside black holes or the initial big bang singularity in the past. Additionally, a poor choice of gauge could also lead to coordinate singularities. In the quantum theory, this behaviour is reflected in the fact that the Hamiltonian constraint (even when it has a precise non-perturbative definition such as in mini-superspace theories) is often not a good Hermitian operator: technically, it fails to be essentially self-adjoint. This means it cannot be exponentiated  to a unitary time-evolution (or  this exponentiation is ambiguous). This is a technical obstruction to directly applying the group-averaging technology. We will encounter such a problem in JT gravity, and give a proposal for circumventing it which gives physically sensible results.

\subsection{Outline}

We now summarise the remainder of the paper, highlighting a few new results for dS JT gravity.

\begin{figure}
    \centering
        \includegraphics[width=5in]{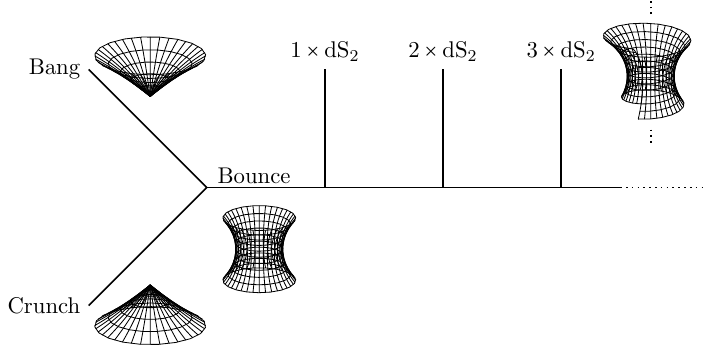}
    \caption{The `fishbone' is the space of connected closed universe spacetimes with constant positive curvature $\mathcal{R}=2$. This divides into sectors, each of which has a single real parameter labelling different solutions. On the left are big bang and big crunch solutions with an initial or final Milne singularity, which we call the \textbf{bang} and \textbf{crunch} sectors. The `spine' of the fishbone gives \textbf{bounce} solutions which expand exponentially in both past and future, parameterised by the minimum radius of the universe. Finally, the `ribs' are an infinite sequence of $\mathbf{n\times\text{\textbf{dS}}_2}$ sectors labelled by positive integers $n$ (joining the `bounce' branch at radius $n$). These spacetimes are $n$-fold covers of dS$_2$, except that  the identification of the angular periodic spatial coordinate comes along with a boost (or static patch time translation).
    The solutions of JT are described by one more real parameter which gives the magnitude of the dilaton solution, which is the conjugate momentum to the parameter describing the geometry. The phase space is thus the cotangent space of this fishbone (at least away from degenerate geometries, which are more subtle).\label{fig:fishbone}}
\end{figure}

In \textbf{\autoref{sec:phasespace}} we discuss the classical theory, in particular the phase space of (connected) closed universes. This part of the paper has significant overlap with \cite{Alonso-Monsalve:2024oii}. The first step is a classification of all possible geometries up to diffeomorphism, which is simple because the dilaton equation of motion imposes constant curvature $\mathcal{R}=2$. The resulting space of geometries is the `fishbone' pictured in figure \ref{fig:fishbone} and explained more in the caption, which is locally one-dimensional but with various qualitatively different  sectors joining at vertices.

The phase space is completed by finding solutions for the dilaton $\Phi$.  For a given geometry (away from the vertices) there is a one-dimensional space of solutions, giving us a two-dimensional phase space. By computing Peierls brackets, we determine that the magnitude of the dilaton is canonically conjugate to a natural coordinate on the fishbone labelling the geometries. The result is that the phase space is the cotangent bundle of the fishbone (with some subtleties at the vertices). To make later contact with the canonical quantum theory, we also analyse this phase space in the Hamiltonian formalism as an extended phase space where we impose Hamiltonian and momentum constraints $\mathcal{H}(\theta)=\mathcal{P}(\theta)=0$ (depending on a spatial coordinate $\theta$).

When we turn to the quantum theory, the technical aspects are greatly simplified by the existence of an \emph{exact} mini-superspace description, which we introduce in \textbf{\autoref{sec:MSS}}. This reduces the above constraints to a single Hamiltonian constraint $H=0$ from the total energy (not the full Hamiltonian density $\mathcal{H}(\theta)$). We describe this theory as an `unconstrained' quantum mechanics with two degrees of freedom --- the dilaton $\Phi(t)$ and the size of the universe $a(t)$ --- along with a constraint operator $H$: much of the rest of the paper is concerned with the details of imposing $H=0$. Because of big bang/crunch singularities, $H$ is not essentially self-adjoint, and we motivate a prescription for dealing with this. Finally, we describe the algebra of gauge-invariant observables, which is naturally described as an $\mathfrak{sl}(2,\RR)$ algebra with generators $G_1,G_2,G_3$ (which commute with $H$), and a single relation from fixing the Casimir.

Such a mini-superspace has been previously used, though we clarify the precise sense in which it is exact (which differs from some previous claims \cite{Iliesiu:2020zld,Maldacena:2019cbz}). Specifically, the mini-superspace theory is exact in a gauge with Cauchy surfaces of constant extrinsic curvature (\emph{not} constant dilaton $\Phi$). This means that the mini-superspace dilaton corresponds in the full theory to the zero-mode (spatial average) $\Phi_0(t)=\int \frac{d\theta}{2\pi} \Phi(t,\theta)$. However, while this mini-superspace gives an exact description of the bang, crunch and bounce sectors it entirely fails to capture the additional $n\times \mathrm{dS}_2$ sectors giving the `ribs' of the fishbone of figure \ref{fig:fishbone} (since these spacetimes have no Cauchy surface of constant extrinsic curvature). We do not attempt to describe the quantum theory in these other sectors in detail.

In \textbf{\autoref{sec:WDW}} we begin our discussion of the physical Hilbert space by describing solutions of the Wheeler-DeWitt equation $H|\psi\rangle=0$ (the invariant states in the above terminology). We do this in a variety of different representations and choices of basis. The first does this in a way which takes full advantage of the simplicity of JT gravity (i.e., linearity of the action in the dilaton) by using wavefunctions $\psi(a,k)$, where $k$ is the momentum conjugate to $\Phi$ (corresponding to $a\mathcal{K}$, where $\mathcal{K}$ is the extrinsic curvature). The result makes explicit contact with the classical phase space, showing that solutions correspond to functions on the space of geometries in figure \ref{fig:fishbone}. We then describe wavefunctions which diagonalise each of the observables $G_i$, using both $\psi(a,k)$ and the Fourier-transformed $\psi(a,\Phi)$ representations. We also describe another basis of asymptotic in- and out-states, which correspond to the asymptotic boundary conditions with very large universes, $a\to\infty$ with fixed $\frac{\Phi}{a}$, as is  familiar in JT (in AdS and in previous work in dS). Finally, we give a basis which makes explicit contact with the representation theory of the $\mathfrak{sl}(2,\RR)$ algebra.

While the solutions to the Wheeler-DeWitt equations give us a description of the space of physical states, this is not enough for a Hilbert space: we also need an inner product. In \textbf{\autoref{sec:IP&GA}} we implement the group-averaging procedure to define an inner product. Describing wavefunctions as functions on the space of possible geometries, this inner product turns the space of states into a familiar $L^2$ Hilbert space with a specific measure. It is manifestly positive-definite, and the observables $G_i$ become Hermitian operators on this Hilbert space.

We conclude the study of the mini-superspace in \textbf{\autoref{sec:KG+}} by exploring the role of Klein-Gordon inner products and various generalisations, as sketched above. In particular, we see explicitly that the Klein-Gordon inner product on `positive frequency' solutions agrees with the group-average inner product to all orders in perturbation theory (and negative frequency solutions agree with minus the group-average inner product). Nonetheless, these are not identical non-perturbatively: there is a difference which would scale as $e^{-\frac{1}{\hbar}\#}$ if we restored units to write $\hbar$ explicitly.

We then revisit the full two-dimensional theory in \textbf{\autoref{sec:2D}}, explaining more details about the connection to the mini-superspace theory (now armed with the technology introduced in the earlier sections). In particular, we explain how the `Schwarzian' \cite{Maldacena:2019cbz,Cotler_2020} arises from a functional Fourier transform between dilaton $\Phi$ and its conjugate momentum $k=a\mathcal{K}$. We also  comment on a similar analysis in the $n\times \mathrm{dS}_2$ sectors, where we find that the wavefunctions are computed by a novel Schwarzian theory.

In  \textbf{\autoref{sec:Qmatter}} we sketch the generalisation of coupling the theory to matter.

We conclude with a discussion in \textbf{\autoref{sec:disc}}. In particular we cover the Hartle-Hawking state, closed universes in the AdS theory, and speculations on non-perturbative topology-changing effects.

\section{The classical phase space}\label{sec:phasespace}

We begin by understanding the classical phase space of closed universes in dS JT gravity, with action
\begin{equation}
     I = \frac{1}{4\pi}\int d^2x \sqrt{-g} \Phi (\mathcal{R}-2).
     \label{dS Action}
\end{equation}
The classical phase space is the space of solutions to the equations of motion modulo diffeomorphisms\footnote{Specifically, we gauge diffeomorphisms which preserve orientation and time-orientation. We can take these orientations to be  additional discrete data which come along with the metric $g$.}, with specified boundary conditions.  For our purposes, we require that there are no spatial boundaries. More precisely, we are looking for globally hyperbolic solutions containing a compact initial data Cauchy surface (in particular, singularities in the past or future are not excluded).

There are (at least) two possible approaches to identifying the phase space. For the first `covariant' approach we can directly classify the space of all possible solutions (and calculate the symplectic structure from Peierls brackets). For the second `Hamiltonian' approach we choose some distinguished time coordinate and identify points in phase space with initial data (coordinates and conjugate momenta) at $t=0$; the diffeomorphism invariance gives rise to Hamiltonian and momentum constraints that must be subsequently imposed to define the physical phase space. The first approach is more straightforward for identifying the full phase space (and proving that nothing has been missed). The second approach makes closer contact with the quantum theory (particularly with the `quantise first, constrain later' approach that we advocate). Given these complementary benefits, we examine each in turn.

\subsection{The phase space from quotients of the universal cover}

The equation of motion from varying the dilaton $\Phi$ tells us that classical solutions have constant curvature $\mathcal{R}=2$. This alone is extremely constraining, so we start by identifying the space of allowed constant curvature geometries, leaving an examination of the dilaton solutions for later. In two dimensions, $\mathcal{R}=2$ completely determines the geometry locally, implying that the metric is locally isometric to dS$_2$. There is nonetheless a non-trivial space of solutions globally, given by quotients of  $\widetilde{\mathrm{dS}_2}$, the universal cover of dS$_2$ (or of a subset of this geometry). As will be explained below, the group of (orientation and time-orientation preserving) isometries of $\widetilde{\mathrm{dS}_2}$ is given by $G= \widetilde{SO^+(2,1)}\simeq \widetilde{SL(2,\RR)}$, the universal cover of $SO^+(2,1)\simeq PSL(2,\RR)$ (the $+$ denotes the restriction to the connected component of the identity, preserving time orientation, though we will often omit this notation below). The quotient group is a subgroup of $G$, which is isomorphic to the fundamental group of spacetime. This must be an infinite cyclic group, since a two-dimensional Lorentzian geometry with a compact spatial slice must have topology $S^1\times \RR$. Such a group is generated by a single element, so we can classify inequivalent spacetimes by the conjugacy classes of $G$ (up to inversion).\footnote{For an alternative account following the same idea see \cite{Alonso-Monsalve:2024oii}.}

To proceed we need to know something about $\widetilde{\mathrm{dS}_2}$ and its isometry group $G=\widetilde{SO(1,2)}$. First, dS$_2$ can be represented as the one-sheeted hyperboloid $-T^2+X^2+Y^2=1$ embedded in the three-dimensional Lorentzian spacetime  $\RR^{1,2}$ with metric $ds^2=-dT^2+dX^2+dY^2$. This manifestly has symmetry group $SO(1,2)$, the Lorentz group of the embedding space. The universal cover $\widetilde{\mathrm{dS}_2}$ is obtained by unwrapping the circle direction. Concretely, if we write $T=\sinh t$, $X=\cosh t \cos\theta$ and $Y=\cosh t\sin\theta$ then the induced metric of the hyperboloid  is
\begin{equation}\label{eq:dS2metric}
    ds^2 = -dt^2 + b^2\cosh^2 t \,d\theta^2
\end{equation}
with $b=1$ (we make use of other values of $b$ in a moment). We get $dS_2$ by taking $\theta$ to be periodic with period $2\pi$, while for $\widetilde{\mathrm{dS}_2}$ we simply allow $\theta$ to be real-valued. The isometry group $G$ is similarly obtained from $SO(1,2)$ by unwrapping the compact direction corresponding to rotations. So an element of $G$ can be labelled (locally) by an element of $SO(1,2)$ along with an integer keeping track of the number of rotations by $2\pi$.

The conjugacy classes of $G$ correspond either to rotations or boosts in $SO(1,2)$. Simplest to understand are the rotations, which are all conjugate to translations in $\theta$ acting on $\widetilde{\mathrm{dS}_2}$ (lifting from $SO(1,2)$ to $\widetilde{SO(1,2)}$ has the effect that the conjugacy class is labelled by a real parameter, rather than a $2\pi$ periodic angle of rotation). Choosing the rotation angle to be $2\pi b$, we get `bounce' geometries with metric \eqref{eq:dS2metric}, where  $\theta$ remains a $2\pi$-periodic angle but $b$ is now a free (positive) parameter.

Considering boosts in $SO(1,2)$, the lift to $\widetilde{SO(1,2)}$ has a more important effect, with conjugacy classes labelled by a real boost parameter and an integer number of $2\pi$ rotations. If this integer is zero, the quotient of $\widetilde{\mathrm{dS}_2}$ splits into disconnected pieces. The interesting regions (which give Lorentzian geometries with compact spatial sections) are `Milne patches', which give either big bang or big crunch geometries. To make this explicit, we can choose coordinates $T=\sinh t\cosh (\beta\theta)$, $X=\cosh t$, $Y=\sinh t\sinh (\beta\theta)$ on $\mathrm{dS}_2$ which make boost symmetries manifest: we have included a boost parameter $\beta>0$ so that the boost in question maps $\theta\mapsto\theta+2\pi$. Before the quotient, the region $t>0$ covers the future Milne patch where $X>1,T>0,Y>0$ (while $t<0$ covers a past Milne patch). The metric \eqref{eq:dS2metric} becomes
\begin{equation}\label{eq:MilneMetric}
    ds^2 = -dt^2+ \beta^2 \sinh^2 t \, d\theta^2,
\end{equation}
and the quotient simply identifies $\theta$ as a $2\pi$ periodic angle. The region $t>0$ is a `big bang' geometry, beginning with an initial singularity (a Lorentzian analogue of a conical singularity) at $t=0$. The region $t<0$ (which we regard as a separate solution, not evolving through the singularity\footnote{Some additional motivation comes from including matter, which we comment on below. We will have more to say about this in the quantum theory.}) is a `big crunch' geometry.

Despite appearances, we can in fact continuously pass between the bounce, bang and crunch geometries discussed so far, via a  pair of special isolated points (corresponding to a `null rotation' in $SO(1,2)$) reached by a limit $b\to 0$ or $\beta\to 0$. To see this, it is nice to describe all these solutions at once with the metric
\begin{equation}
    ds^2 = -dt^2 + \left(\frac{c_+ e^t+c_- e^{-t}}{2}\right)^2 d\theta^2
\end{equation}
for constants $c_\pm$, and take $\theta$ to be $2\pi$ periodic (and we restrict to the region where $c_+ e^t+c_- e^{-t}>0$). Note that rescaling $c_+$ by a positive constant $\lambda$ and $c_-$ by $\lambda^{-1}$ gives the same metric if we translate $t$ by $\log\lambda$, so only the product $c_+c_-$ is invariant, giving $b^2$ for the bounce solutions and $-\beta^2$ for bang or crunch. Nonetheless, in this form it is manifest that by fixing $c_+$ and taking $c_-$ from positive to negative, we pass from a bounce solution to a bang solution with smooth changes to initial data at $t=0$ (which remains true when we add a dilaton). Similarly, we can pass from bounce to crunch (though to go from bang to crunch we always must pass through bounces).\footnote{A natural topology on the space of these geometries (the quotient topology on the space of initial data on a constant $t$ slice) here gives rise to a non-Hausdorff space --- the `branching line'.}

The final class of solutions comes from boosts along with non-zero integers $n$ (which we can take to be positive WLOG by going to the inverse if necessary) labelling the number of $2\pi$ rotations when we lift to the universal cover. The resulting geometry is given by taking an $n$-fold cover of dS$_2$ (i.e., the metric \eqref{eq:dS2metric} with $b=n$ and $0<\theta<2\pi$), and gluing the edges $\theta=0,\pi$ with a boost. Concretely, this means identifying a point $(\theta=0,t)$ with $(\theta=2\pi,t+2\pi\beta)$ for some real boost angle $\beta$ (which parameterises possible geometries). For $\beta=0$ we recover the `bounce' geometries for special integer-quantised values $b=n$, but non-zero $\beta$ we find a new branch of solutions. We will refer to this class of spacetimes as $n\times$dS$_2$ geometries.\footnote{We would like to thank Don Marolf for pointing out this class of solutions.}

We summarise this space of possible geometries by a `fish-bone' as shown in figure \ref{fig:fishbone}. The bounce solutions make up the `spine', parameterised by $b>0$. Two branches of a `tail' emanate from the $b=0$ end, from bang and crunch solutions each labelled by $\beta>0$. At each $b=n$ for positive integer $n$, a pair of `ribs' giving $n\times$dS$_2$ geometries attach (labelled by positive $\beta$, joined to the spine at $\beta=0$).


To complete this into the full phase space, we must also consider solutions for the dilaton. As is well-known, the equation of motion from varying the metric tells us that $\nabla\Phi$ is orthogonal to a Killing vector field (so in particular,  $\Phi$ is constant along integral curves of the vector field). This means that we can classify dilaton solutions by classifying symmetries of each of the metrics above. Fortunately, in each case there is a single one-parameter isometry group (with the exception of the isolated points where the `ribs' intersect the `spine'). This means that there is a single dilaton solution up to normalisation. Including this single normalisation parameter gives us (locally away from special points) a two-dimensional phase space of solutions parameterised by one parameter for the geometry, and one for the strength of the dilaton. For a useful way to parameterise this, note that the following quantity is constant when the equations of motion are satisfied:
\begin{equation}\label{eq:Phiconst}
    M=\Phi^2 + g^{ab}\nabla_a \Phi \nabla_b \Phi = \begin{cases}
        \varphi^2 & (\text{when positive}) \\
        -\tilde{\varphi}^2 & (\text{when negative})
    \end{cases} \qquad \text{is constant on-shell.}
\end{equation}
We call this $M$ following \cite{Iliesiu:2020zld,Nanda:2023wne}, following the AdS theory where it corresponds to the mass of a black hole. Using this parameterisation, we can explicitly summarise the geometries and dilaton solutions as follows:
\begin{align}
    \text{Bounce:}\quad & ds^2 = -dt^2 + b^2\cosh^2 t d\theta^2, \quad \Phi = \tilde{\varphi} \sinh t \label{eq:bouncesol} \\
    \text{Bang/crunch:}\quad & ds^2 = -dt^2 + \beta^2\sinh^2 t d\theta^2, \quad \Phi = \varphi \cosh t \qquad (t>0/t<0)  \label{eq:BCsol}\\
    \text{$n\times$dS$_2$:}\quad & ds^2 = -dt^2 + n^2\cosh^2 t d\theta^2, \quad \Phi = \varphi\cosh t  \sin n \theta \quad (\theta=0,t)\sim (\theta=2\pi,t+2\pi \beta)\label{eq:nAdSsol}
\end{align}
For generic geometries (away from the vertices of figure \ref{fig:fishbone}), each different real value of $\varphi$ or $\tilde{\varphi}$ is a different solution (though at special points there are additional symmetries which lead to additional identifications, for example at  $\beta=0$ in the last case we have $\varphi\sim-\varphi$).

Each solution has a single-parameter family of symmetries. For bounce, bang and crunch solutions these are the manifest rotational symmetries generated by $\partial_\theta$. For the $n\times\mathrm{dS}_2$ solutions, we have a more complicated symmetry generated by $\cos n\theta \partial_t - \frac{1}{n}\sin n\theta \tanh t \partial_\theta$. In the `static patch' causally related to the worldline $\theta=0$ this is a timelike Killing vector so we have a static solution in this region (and in fact this holds for $n$ different patches where we take $\theta$ to be a multiple of $\frac{2\pi}{n}$).

The way in which different sectors of phase space are patched together (between bang/crunch and bounce solutions at $b=0$ or $\beta=0$, and at $b=n$) is somewhat subtle. We do not work out all the details here, though we will confront related issues in the quantum theory later, at least for the bang/crunch/bounce vertex.

\begin{figure}
    \centering
    \includegraphics[width=7in]{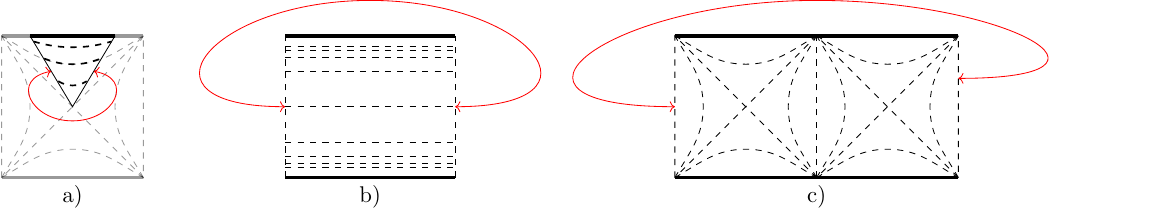}
    \caption{Each spacetime is a piece of $\widetilde{\mathrm{dS}_2}$, with some identifications. These are illustrated here, along with integral curves of the one-parameter isometry group (which are also the curves of constant dilaton) in each case. \textbf{a)} Identifying two curves related by a boost in the future Milne wedge of one copy of global $\mathrm{dS}_2$ lead to a big bang geometry. A similar identification in the past Milne wedge would yield a big crunch geometry. \textbf{b)} Identifying the worldlines of two static observers at different spatial locations. These result in bounce geometries with the minimal spatial volume depending on the minimal proper separation between the two worldliness.  \textbf{c)} Finally, if the worldines in the previous case are separated by exactly an integer multiple of $2\pi$ in $\theta$, we can perform the identifications with a relative boost between the two worldines. This results in what we call the $n\times\mathrm{dS}_2$ spacetimes.}
    \label{fig:enter-label}
\end{figure}

\subsection{Symplectic structure: Peierls brackets}

Having identified the space of solutions, we turn it into a phase space by determining the symplectic structure (Poisson brackets). In the covariant language we have been using so far, it is natural to do this using Peierls brackets. This is somewhat outside the main line of development we subsequently follow, so we will give a brief and incomplete account, assuming prior knowledge of the relevant definitions (see \cite{Peierls:1952cb,Marolf_1994} for background).

To compute Peierls brackets, it is useful to write functions on phase space as gauge-invariant integrals, which we can consider adding as deformations to the action. These integrals should be compactly supported in time (at least for the class of solutions under consideration) so that before and after the deformation turns on, the solutions coincide with those of the undeformed theory. One class of candidate examples is given by
\begin{equation}
    I_F = \frac{1}{2\pi}\int d^2x\sqrt{-g} F(\Phi),
\end{equation}
where $F$ is a compactly supported function of the dilaton profile. For the integrand to vanish at early and late times (as required for computing Peierls brackets), we need the  dilaton at late and early times to take values outside the support  of $F$. This will apply to bounce solutions for any compactly supported $F$, and to bang/crunch solutions for which $\varphi$ (the value of $\Phi$ at the final/initial singularity) lies outside the support of $F$. It does not apply to the $n\times$dS$_2$ solutions with spacelike dilaton, because the range of $\Phi$ on any Cauchy surface in the distant past/future approaches the full real line: we will have to do something else in that case. 

We first identify $I_F$ with a function on phase space (as a function of $(b,\tilde{\varphi})$ or $(\beta,\varphi)$. The result in the relevant cases is
\begin{equation}\label{eq:IFvalue}
    I_F =  \int_{-\infty}^\infty F(\Phi)d\Phi \times \begin{cases} \frac{b}{|\tilde{\varphi}|} & \text{(bounce)} \\\frac{\beta}{|\varphi|} &   \text{(bang/crunch)},
    \end{cases}
\end{equation}
which depends on $F$ only through the constant prefactor $\int F $.


Now to compute the Peierls brackets of $I_F$ with any quantity, we identify the change in the solution when we perturb the action by adding $\varepsilon I_F$ to linear order in $\varepsilon$. The equation of motion from varying the dilaton becomes $\mathcal{R} = 2+2\varepsilon F(\Phi)$. We can solve this (by variation of parameters) to first order perturbing around a solution, fixing the gauge $ds^2 = -dt^2+a(t)^2 d\theta^2$ (using rotational symmetry), and demanding that the variation $\delta a$ vanishes in the past (when $\Phi$ lies outside the support of $F$). In the future, the solution for $\delta a$ can be written as a linear combination of $\sinh t$ and $\cosh t$. One of these ($\sinh t$ for bounce solutions and $\cosh t$ for bang/crunch) is a linearised gauge transformation of the original solution (a shift in $t$), so does not represent a physical change. The other is a physical perturbation to the solution giving the variation in $b$ or $\beta$, and so its coefficient is (by definition) the Peierls bracket $\{I_F,b\}$ or $\{I_F,\beta\}$. Taking the bounce solutions as an example, we find
\begin{equation}
    \{I_F, b\} = -\sgn \tilde{\varphi} \frac{b}{\tilde{\varphi}^2} \int F(\Phi)d\Phi.
\end{equation}
As a check, we find that this is indeed proportional to the same integral of $F$ that appeared before, and using \eqref{eq:IFvalue} we get
\begin{equation}\label{eq:phibPoisson}
    \{\tilde{\varphi},b\}=1,
\end{equation}
so that $b$ and $\tilde{\varphi}$ are canonically conjugate variables. We can check this result independently from the deformation to the dilaton solution (giving us $\{I_F,\tilde{\varphi}\}$) and by using different deformations of the action. For example,   $\int d^2x\sqrt{-g} (\Phi^2+(\nabla\Phi)^2)F(\Phi)$ gives us observables proportional to $b|\tilde{\varphi}|$ using \eqref{eq:Phiconst}. Similar results apply in the bang/crunch sectors. 

For the $n\times$dS$_2$ solutions, it is somewhat trickier to identify a covariant expression for phase space quantities because integrals of local quantities like $I_F$ are never compactly supported due to the time-translation symmetries in the static patch. One possibility to define $\varphi$ is to use the value of $|\Phi|$ at the horizons, which is clearly compactly supported and gauge-invariant. To compute Peierls brackets, we can use the integral
\begin{equation}
    \int d^2 x \det(\nabla_a\nabla_b \Phi) \delta^{(2)}(\nabla_a\Phi) F(\Phi)
\end{equation}
for any function $F$, which localises to the horizons (giving simply the sum of $F(\Phi)$ at each horizon,  with signs from the determinant). The determinant of the Hessian factor is included to make this coordinate-independent. Extrapolating from the previous results, the expectation is that the dilaton scaling $\varphi$ is conjugate to the boost angle $\beta$, though we have not checked this.

See \cite{Alonso-Monsalve:2024oii} for a rather different approach to calculate the symplectic form.

\subsection{Hamiltonian analysis of phase space}
\label{ssec:hamAnalysis}

For a complementary perspective, we can also examine the phase space from a `Hamiltonian' perspective, where solutions are described by initial data after taking constraints into account.

The starting point is to parameterise the metric in ADM form,
\begin{equation}
    ds^2 = -N^2 dt^2 + a^2 (d\theta + N_\perp dt)^2,
\end{equation}
where $N$ and $N_\perp$ are the usual lapse and shift, and we have written the one-dimensional spatial metric as $ds^2=a^2 d\theta^2$ (with $\theta$ a $2\pi$-periodic angular coordinate). Putting this decomposition into the action \eqref{dS Action} and integrating by parts gives
\begin{equation}
    I = -\int dt \frac{d\theta}{2\pi} \left[N\left(\frac{\Phi''}{a} -\frac{a'\Phi'}{a^2}+ a \Phi\right) + N^{-1}\left(\dot{\Phi}-N_\perp \Phi'\right)\left(\dot{a}-(N_\perp a)'\right)  \right].
    \label{eq:ADMAction}
\end{equation}
Throughout, we use dots $\dot{\phantom{a}}$  and primes $'$ to denote differentiation with respect to $t$ and $\theta$ respectively. From  \eqref{eq:ADMAction}, we can read off the conjugate momenta to $a$ and $\Phi$ to be
\begin{align}
    p &= \pi_a = -N^{-1}\left(\dot{\Phi}-N_\perp \Phi'\right), \\
    k &= \pi_\Phi = -N^{-1}\left(\dot{a}-(N_\perp a)'\right).
\end{align}
These are the derivative of $\Phi$ with respect to a unit normal (pointing in the past direction), and $a$ times the extrinsic curvature respectively. We have normalised these to give Poisson brackets $\{a(\theta),p(\theta')\} = 2\pi \delta(\theta-\theta')$ and $\{\Phi(\theta),k(\theta')\} = 2\pi \delta(\theta-\theta')$. We can perform a Legendre transform to get the first-order action
\begin{equation}
    I = \int dt \frac{d\theta}{2\pi}\left[p \dot{a} + k \dot{\Phi} - N \mathcal{H} - N_\perp \mathcal{P} \right],
\end{equation}
where the Hamiltonian and momentum constraints are
\begin{align}
    \mathcal{H} &= -k p + a \Phi - \frac{a'\Phi'}{a^2}+\frac{\Phi''}{a}, \\
    \mathcal{P} &= k \Phi'- a p',
\end{align}
where everything is a function of $\theta$.

Since the lapse and shift $N$, $N_\perp$ appear without derivatives, they give rise to constraints $\mathcal{H}(\theta)=\mathcal{P}(\theta)=0$. Before imposing constraints we have a large `unconstrained phase space' parameterised by $a,\Phi$ and their conjugate momenta $p,k$, each of which is a function of $\theta$. To get from this to the physical phase space we restrict to solutions to the constraints, and also identify configurations related by equivalence under gauge transformations generated by the diffeomorphisms corresponding to the vector field $\xi(\theta)\partial_\theta$, and $\mathcal{H}[\eta]$ similarly generates time translations where the Cauchy surface moves at $\theta$-dependent speed $\eta(\theta)$ (with respect to a unit normal vector). The algebra is
\begin{equation}\label{eq:algebra}
    \{\mathcal{P}[\xi_1],\mathcal{P}[\xi_2]\} = \mathcal{P}[\xi_1\xi_2'-\xi_1 \xi_2'], \quad  \{\mathcal{P}[\xi],\mathcal{H}[\eta]\} = \mathcal{H}[\xi\eta'], \quad \{\mathcal{H}[\eta_1],\mathcal{H}[\eta_2]\} = \mathcal{P}\left[\frac{\eta_1\eta_2'-\eta_1 \eta_2'}{a^2}\right].
\end{equation}
The first is the algebra of one-dimensional diffeomorphisms (the Lie bracket of vector fields). The second says that $\eta$ is a scalar in terms of the spatial manifold. For the final equation, note that the Poisson bracket of two Hamiltonian constraints depends explicitly on $a$, so the constraints do not form a Lie algebra.

The constant \eqref{eq:Phiconst} will also be useful, so we note its expression in these canonical variables:
\begin{equation}\label{eq:Phiconst2}
    M=\Phi^2+(\nabla\Phi)^2 = \Phi^2+ \frac{(\Phi')^2}{a^2} - p^2.
\end{equation}
The constancy of this quantity is reflected in the fact that it has vanishing Poisson bracket with the constraints when the constraints are satisfied. Explicitly, we have
\begin{equation}
    \begin{aligned}
        \{\mathcal{P}[\xi],M\} &= -\frac{2\xi}{a}\left(p\mathcal{P} + \Phi' \mathcal{H}\right) \\
        \{\mathcal{H}[\xi],M\} &= -\frac{2\xi \Phi'}{a^3} \mathcal{P},
    \end{aligned}
\end{equation}
where the right hand side is evaluated at the same value of $\theta$ chosen to define $M$ on the left.

\subsubsection{Gauge fixing}

To analyse the phase space in practice, it is convenient to impose some gauge-fixing conditions to remove the redundancy. The danger is that we may not find a good gauge conditions globally, so either some physical solutions never obey the chosen gauge conditions, or can obey the conditions in multiple ways. Nonetheless, we can always fall back on the above analysis to check such issues.

First, gauge-fixing the momentum constraints (i.e., diffeomorphisms of our one-dimensional spatial manifold at fixed time) is relatively straightforward: we can demand that $\theta$ is proportional to proper length with the condition $a'(\theta)=0$ (so $a$ is a constant equal to the proper length of the circle divide by $2\pi$). This is certainly a good gauge condition, except that it leaves a residual gauge symmetry of rigid translations, generated by total momentum $P=\int \frac{d\theta}{2\pi} \mathcal{P}(\theta)$ (this has vanishing Poisson bracket with the gauge condition $a'$).

It is more challenging to find a good gauge to fix the Hamiltonian constraints, which works universally for all states. We first use the gauge condition $k'=0$, which fixes the initial Cauchy surface to be a surface of constant extrinsic curvature. This is a useful generalisation of the $k=0$ gauge fixing to a geodesic. From the solutions we found above, we know that this will be sufficient to describe the bang, bounce and crunch states by fixing to constant $t$ slices. However, it is not obvious whether this gauge  captures the $n\times$dS$_2$ solutions with nonzero time shift $\beta$, and in fact it turns out not to: these spacetimes do not contain any closed geodesics, or any closed curves of constant extrinsic curvature! We will come back to these later.

Fixing $k'=a'=0$, the momentum constraint then implies that $p-\frac{k}{a}\Phi$ is a constant. The Hamiltonian constraint in turn tells us that $\Phi'' + (a^2-k^2)\Phi$ is a constant. This has non-constant solutions for $\Phi$ only if $a^2-k^2=n^2$ is an integer. With the exception of this special case (which we will return to in a moment), the conclusion is that fixing $a$ and $k$ to be independent of $\theta$ implies the same for $p$ and $\Phi$, and these constant values are constrained by $a\Phi-k p=0$. We can also look for residual diffeomorphisms by finding the  most general linear combination of constraints which has vanishing Poisson brackets with both $a'$ and $k'$, at a point where both the constraints and gauge conditions are satisfied. We find rigid spatial translations generated by $P$ as above, as well as rigid time translation generated by total Hamiltonian $H= \int \frac{d\theta}{2\pi} \mathcal{H}(\theta)$. Additional residual gauge symmetries appear for the exceptional $k^2-p^2=n^2$ case only. These cases correspond to the $n\times$dS$_2$ geometries with no time-shift $\beta=0$: the additional residual gauge symmetries are a result of the $\mathfrak{so}(2,1)$ isometries of these spacetimes. While we can describe dilaton profiles with spacelike variation (as in \eqref{eq:nAdSsol}) in this gauge, there is no freedom to turn on non-zero time-shift $\beta$ so we cannot describe the `ribs' in the fishbone diagram \ref{fig:fishbone} using this gauge.

The upshot with the $a'=k'=0$ gauge condition is that the original infinite-dimensional phase space (where variables are functions of $\theta$) is reduced to a finite dimensional phase space parameterised by constant $a,\Phi,p,k$, with a single Hamiltonian constraint $H=a\Phi-k p$. This is a `mini-superspace' truncation of JT gravity, and we will spend most of our time studying the corresponding mini-superspace quantum theory. It captures the bang, crunch and bounce sectors of the full theory, and we will later argue that the mini-superspace theory is exact for such states in the quantum theory (with the appropriate interpretation). Its only failure is that it does not include the $n\times$dS$_2$ sectors with spacelike dilaton variation.


\subsubsection{A gauge for $n\times$dS$_2$ solutions}

To capture this last class of solutions, various obvious alternative gauge conditions are insufficient. We have not found a gauge that makes the analysis as straightforward for the other sectors. While the physics of these solutions is potentially interesting (particularly once we add matter), we will not pursue them much further here, since our main aim in this paper is to illustrate conceptual ideas in canonical quantisation of gravity.

Nonetheless, we briefly summarise one possible gauge which captures these solutions. This gauge is motivated by selecting a Cauchy surface by minimising some functional on the surface (or at least finding a stationary point). The simplest possible functional is the length $\int a d\theta$, but asking for stationary points gives us the gauge $k=0$ which we already saw does not work: the length gets pushed to zero on nearly-null surfaces. We instead try the next simplest, the integral of the square of the extrinsic curvature $\int \frac{d\theta}{2\pi} \frac{k^2}{a}$. This is a covariant expression with respect to one-dimensional diffeomorphisms ($\frac{k}{a}$ is the extrinsic curvature, a scalar, and $a d\theta $ is the volume form), so it has vanishing Poisson bracket with momentum constraints $\mathcal{P}$. Our gauge condition comes from setting its Poisson bracket with Hamiltonian constraints to zero:
\begin{equation}\label{eq:ksqGF}
    \left\{ \mathcal{H},\int \frac{d\theta}{2\pi} \frac{k^2}{a}\right\} = \frac{2}{a^2}\left(k'' + a^2 k - \tfrac{1}{2}k^3\right)=0,
\end{equation}
where we have also fixed the gauge $a'=0$ for spatial diffs.

This gauge condition for $k$ is the equation of motion for a particle moving in a quartic potential $V(k)= \frac{a^2}{2} k^2-\frac{1}{8}k^4$. We require $k(\theta)$ to be $2\pi$ periodic, so we are looking for a solution with period $2\pi$. This can be achieved either by a solution with constant $k$, or by a solution with minimal period $\frac{2\pi}{n}$ for any integer $n\geq 1$ traversed $n$ times. For the constant solutions we have three options for any $a$: $k=0$ and $k=\pm \sqrt{2}a$. These give the bounce, bang and crunch geometries which were already visible in the $k'=0$ gauge (though now on a fixed Cauchy surface with no residual time-translation gauge symmetry). To understand the non-constant solutions, we can examine how the minimal period of a solution changes as we increase the amplitude of $k$ holding $a$ fixed. For small amplitude, the period approaches $\frac{2\pi}{a}$ (from simple harmonic motion around the quadratic minimum), and as the amplitude increases the period increases monotonically, going to infinity as the maximal value of $|k|$ approaches the maxima of the potential. This means that we have a unique solution of minimal period $\frac{2\pi}{n}$ when $a>n$: these correspond to the $n\times$dS$_2$  solutions, joining to the constant $k$ bounce solutions at $a=n$ when their amplitude goes to zero. The upshot is that the space of solutions to the gauge condition \eqref{eq:ksqGF} precisely aligns with the fishbone in figure \ref{fig:fishbone}.

\begin{figure}
    \centering
    \includegraphics[width=0.65\linewidth]{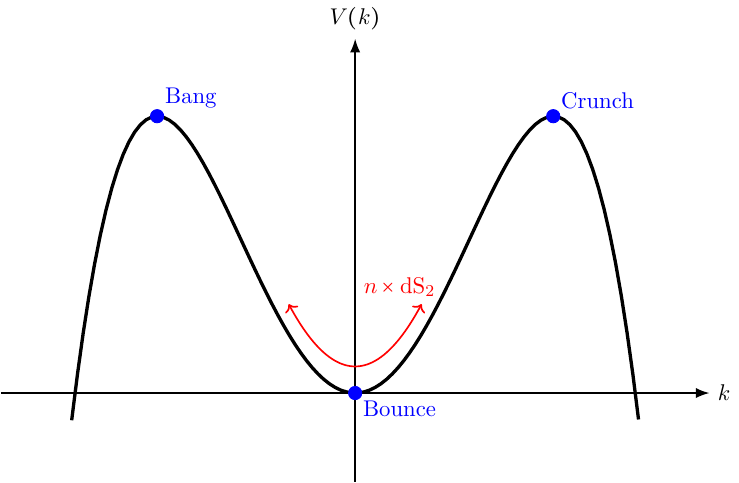}
    \caption{The gauge condition \eqref{eq:ksqGF}  is the equation of motion for a particle moving in a quartic potential $V(k)$ depicted here, where we require solutions to be $2\pi$ periodic. Solutions with $k(\theta)=\mathrm{constant}$ correspond to a particle sitting at the extrema of this potential with $k(\theta)= 0,\pm\sqrt{2a}$ corresponding to bouncing, crunching, and banging geometries respectively. Non-constant $k(\theta)$ satisfying the gauge condition have solutions which oscillate about the minima of the potential with minimal period $\frac{2\pi}{n}$, and correspond to $n\times\mathrm{dS}_2$ geometries not captured using the $k'=0$ gauge.}
    \label{fig:kPotential}
\end{figure}
To complete the phase space we look for dilaton solutions, $\Phi(\theta)$ and $p(\theta)$ which satisfy the constraints. To get straight to the solutions, we can note that the value of the functional $\int \frac{d\theta}{2\pi} \frac{k^2}{a}$ gives us a function on the physical phase space, which generates a family of dilaton solutions. Concretely, we have $\delta \Phi = \{\Phi,\int \frac{d\theta}{2\pi} \frac{k^2}{a}\}$ and similarly for $p$. This infinitesimal deformation immediately gives us finite solutions because the constraints are linear in $\Phi,p$:
\begin{equation}
    \Phi = \lambda \frac{k}{a}, \qquad p = \lambda \frac{k^2}{2a^2}.
\end{equation}
To compare this to the earlier solutions, we can use the constant $\varphi^2$ as given in \eqref{eq:Phiconst} (which has the interpretation as $\pm$ the value of dilaton on the horizons of the static patches).  From \eqref{eq:Phiconst2} we get $\varphi^2  = \frac{8\mathcal{E}}{a^4}\lambda^2$, where $\mathcal{E}= \frac{1}{2}(k')^2 + \frac{a^2}{2} k^2-\frac{1}{8}k^4$ is the `energy' of the periodic solution $k(\theta)$ to the constraints. We also immediately get the symplectic structure (without needing to use Dirac/Peierls brackets) since $\lambda$ and  $\int\frac{d\theta}{2\pi} \frac{k^2}{a}$ are canonically conjugate by construction.

\subsection{Coupling to matter}
\label{ssec:clMatter}

Most of our above considerations generalise straightforwardly when we couple this theory to matter. For this we can take the matter to be a general two-dimensional field theory, which we couple to the metric but not the dilaton. This standard choice is natural if JT is thought of as a near-extremal reduction  from higher dimensions since coupling to the dilaton is suppressed in that limit.

Since matter does not couple to the dilaton, the equations of motion $\mathcal{R}=2$ from varying $\Phi$ are unchanged. This means that the discussion of the geometry is unaltered. So, our phase space is still described by the fishbone geometry of figure \ref{fig:fishbone} along with additional data about the matter and dilaton.

Since the matter action is independent of the dilaton, the solutions to the matter equation of motion depend on the  metric only. So, a point in the phase space of the matter theory on the geometry in question --- for example, initial data on some arbitrary Cauchy surface --- gives us a solution for the matter fields. Finally, we determine the dilaton solutions. This is modified by the matter because the equation of motion from varying the metric picks up a source from the stress-energy of the matter:
\begin{equation}\label{eq:EOMmatter}
      g_{ab}(\Phi + \nabla^2\Phi)-\nabla_a \nabla_b \Phi  =2\pi  T_{ab}.
\end{equation}
Now, this equation will not have a solution for arbitrary conserved source $T_{ab}$ (obeying $\nabla^aT_{ab}=0$). The reason is that every metric solution has a one-parameter family of continuous symmetries, which are residual gauge transformations leading to additional constraints. For the bang, crunch and bounce sectors this residual symmetry is simply rotation, so we require that the total momentum of matter is zero. Explicitly, the $t\theta$ component of the left hand side of \eqref{eq:EOMmatter} is a derivative with respect to $\theta$, so $P_\mathrm{matter} = \int d\theta T_{t\theta}=0$. For the $n\times$dS$_2$ sectors the residual symmetry is the time-translation of the static patch, so the total static patch Hamiltonian of the matter must vanish. There are a total of $2n$ static patches, which are causal diamonds where the Killing field is timelike, and it is directed to the future and the past in alternating patches; thus, the constraint is that the total energy in `even' static patches equals the total energy in `odd' patches.

In addition to this constraint on the matter state we should also impose equivalence under the residual gauge symmetry, so two solutions which differ by a gauge transformation correspond to the same point in phase space.  For bang/crunch/bounce sectors, this means that two matter configurations which differ by a rotation are the same point in phase space. The upshot is that for a given geometry, we have a matter sector of the phase space given by the usual phase space of matter on that geometry, except that we impose the single constraint corresponding to the symmetry of the metric in question.

Finally, for a given matter state satisfying the above constraints we still have a single degree of freedom in choosing the dilaton solution: we can add any multiple of the homogeneous solution as given in \eqref{eq:bouncesol}, \eqref{eq:BCsol} or \eqref{eq:nAdSsol}. This free dilaton solution is the canonical conjugate of the single metric degree of freedom, as was the case for pure JT.

The upshot is that the phase space with matter is a bundle over the pure dS JT phase space, where the fibre over a given point is given by the phase space of the matter on the corresponding geometry after gauging the single residual symmetry.\footnote{There could be more to say about the intersection between different branches of the fishbone, particularly the $a=n$ points, though we will not worry about these degenerate lower-dimensional surfaces in phase space.} We will comment on the quantum analogue of this statement in section \ref{sec:Qmatter}, though mostly we will study pure dS JT.




\section{The mini-superspace theory}\label{sec:MSS}

\begin{figure}
    \centering
    \includegraphics[width=5in]{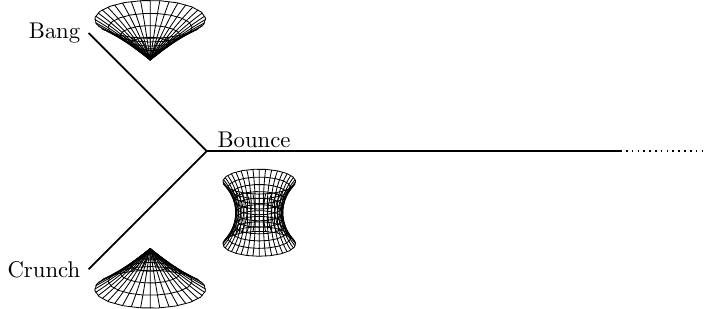}
    \caption{When we impose the $a'=k'=0$ gauge fixing and pass to the mini-superspace theory, we  do not pick up the $n\times\mathrm{dS}_2$ geometries. The space of remaining $\mathcal{R}=2$ geometries is depicted above. There are three remaining sectors --- bouncing, banging, and crunching solutions, each labelled by a single parameter --- meeting at a vertex (which encompasses two distinct geometries).}
    \label{fig:miniPhaseSpace}
\end{figure}

From the classical analysis, we have seen that most of the phase space is correctly described by a mini-superspace model with finitely many degrees of freedom, by a partial gauge-fixing to Cauchy surfaces of constant extrinsic curvature, with a single residual constraint $H=0$.\footnote{Rigid rotations generated by $P$ are also residual gauge symmetries, but $P$ is identically zero when $a'=k'=0$ so this does not impose any additional constraint. This changes if we couple to matter, as described above in section \ref{ssec:clMatter} and below in section \ref{sec:Qmatter}.} These degrees of freedom are $a$ (the volume of the spatial slice over $2\pi$), and dilaton $\Phi$, along with their conjugate momenta $p$, $k$ respectively. The Hamiltonian constraint is
\begin{equation}
    H=-kp+a\Phi.
\end{equation}
Note that we can take $a$ to be positive to describe all distinct spatial geometries, so we restrict to $a>0$; this will be the source of some subtleties in the quantum theory.

As we will later argue in section \ref{sec:2D}, the mini-superspace truncation in fact remains exact in the quantum theory for the class of states that it describes. This exact description should be interpreted in the gauge $k'=0$, with Cauchy surfaces of constant extrinsic curvature. Since $k$ is the momentum conjugate to the dilaton $\Phi$, this gauge-fixing means that the spatially varying modes of $\Phi$ are completely uncertain (an equal superposition over all possible values). The dilaton $\Phi(t)$ in the mini-superspace model should be interpreted as the remaining zero-mode $\Phi(t) = \int \frac{d\theta}{2\pi}\Phi(t,\theta)$ of the two-dimensional theory.

The classical phase space of the mini-superspace model can be lifted straight from the previous section, simply by discarding the $n\times$dS$_2$ sectors that are not captured. Thee fishbone of figure \ref{fig:fishbone} is reduced to the `Y-shaped' geometry of figure \ref{fig:miniPhaseSpace}. The phase space is the cotangent space of this Y geometry (joined at the vertex as before).

\subsection{Quantizing the unconstrained theory}

The main topic of the remainder of the paper is the quantisation of this mini-superspace theory, illustrating ideas of more general applicability. We advocate for a `quantise first, constrain later' approach. This means we start with a quantum theory of all variables (here $a,\Phi$ and momenta $p,k$) and an algebra of constraint operators represented on the corresponding `unconstrained' Hilbert space $\hilb_0$ (here just the Hamiltonian operator $H$). We then impose these quantum constraints to build a physical Hilbert space $\hilb$ from $\hilb_0$, consisting of both a space of physical states and an inner product. Our primary aim is to explain and implement a recipe for this construction.

The alternative approach is to impose constraints first to arrive at the physical classical (symplectic quotient) phase space, and then quantise. While this would be adequate for the present example, more generally it has drawbacks. One is that the `constrain first' approach can eliminate certain important and interesting (usually non-perturbative) quantum effects (examples are given in \cite{Held:2025mai}). Another is that it may not be possible to describe the symplectic quotient phase space in a way that manifestly preserves all symmetries. The way we have presented things so far can be thought of as a hybrid approach, where we impose some constraints classically (by gauge-fixing to $k'=0$), but leave the residual constraint $H=0$ to be imposed after quantisation. After introducing the main ideas in the simpler mini-superspace context, we will comment on a more complete `quantise first' construction in section \ref{sec:2D}.

With this preamble, we can introduce the quantisation of the unconstrained system. With just two degrees of freedom, this is extremely simple: $\hilb_0$ can be described simply by square integrable wavefunctions of the `position' variables $\psi(a,\Phi)$ (defined for $a>0$ and $\Phi$ real). The momentum operators act in the usual way as $p=-i \frac{\partial}{\partial a} $ and $k=-i \frac{\partial}{\partial \Phi}$. The Hamiltonian acts on the wavefunctions as
\begin{equation}\label{eq:Haphi}
    H = \frac{\partial^2}{\partial\Phi\partial a} + a\Phi.
\end{equation}
We can think of this as a Laplacian plus potential $-\frac{1}{2}\nabla^2+V$ on a two-dimensional target space with Lorentzian metric $ds^2 = -2d\Phi da$ (so $a,\Phi$ are `lightcone coordinates'), and potential $V=a\Phi$. This structure of the Hamiltonian constraint is typical in more general theories of gravity.

It will also be useful to use an alternative $a,k$ representation, where wavefunctions are related to the $(a,\Phi)$ representation by Fourier transform,
\begin{equation}
    \psi(a,k) = \int \frac{d\Phi}{\sqrt{2\pi}} e^{-ik\Phi}\psi(a,\Phi), \qquad \psi(a,\Phi) = \int \frac{dk}{\sqrt{2\pi}} e^{ik\Phi}\psi(a,k).
\end{equation}
In this representation, the Hamiltonian is 
\begin{figure}
    \centering
    \includegraphics[width=0.4\linewidth]{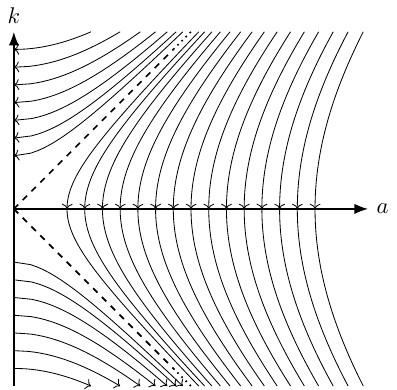}
    \caption{When expressed in the basis of $a$ and $k$ eigenstates the Hamiltonian becomes a first order differential operator that we can interpret as generating boosts on the right half of a two-dimensional Minkowski space with $a$ and $k$ acting as space and time respectively. Each orbit of the boost corresponds to a definite spacetime geometry (a point on the configuration space in figure \ref{fig:miniPhaseSpace}). `Timelike' orbits in the `Rindler wedge' $a>|k|$ correspond to bounce states while orbits in the `Milne' wedges correspond to bang and crunch states.}
    \label{fig:dSflow}
\end{figure}
\begin{equation}\label{eq:Hak}
    H = i\left(k\frac{\partial}{\partial a} +a \frac{\partial}{\partial k}\right).
\end{equation}
This is the generator of a boost in a Minkowski half-space (where we can think of $a>0$ as a spatial coordinate and $k$ as `time'). Since this $H$ is a first-order differential operator, it makes many calculations very simple. In particular, time evolution (i.e., the solution $\psi(t)= e^{-i H t}\psi$ to the differential equation $H\psi(t) = i \partial_t\psi(t)$ with with a given initial state at $t=0$) is
\begin{equation}\label{eq:akevolution}
    e^{-i H t}\psi (a,k) = \psi(a\cosh t+k \sinh t,k\cosh t+a \sinh t)
\end{equation}
(where for now we are ignoring subtleties from the $a>0$ restriction).

We can make this simplicity very explicit by choice of coordinates. We can separate the $(a,k)$ space into three regions: a  `Rindler wedge' $a>|k|$ and two (half) `Milne wedges' $0<a<k$ and $0<a<-k$. In the Rindler region we can define coordinates $b,u$
\begin{equation}\label{eq:Rindler}
    a=b \cosh u,\quad k = b \sinh u \implies H= i\frac{\partial}{\partial u},
\end{equation}
with $b>0$, so the Hamiltonian simply generates translations in $u$. Similarly, in the Milne wedges we use coordinates $\beta,v_\pm$
\begin{equation}\label{eq:Milne}
    a=\pm b \sinh v_\pm, \quad k=\pm \beta \cosh v_\pm \implies H= i\frac{\partial}{\partial v_\pm},
\end{equation}
with $\beta>0$, as well as $v_+>0$ and $v_-<0$ from the restriction to $a>0$.

We can also determine how wavefunctions evolve in the $a,\Phi$ basis by taking the Fourier transform of \eqref{eq:akevolution} with delta-function wavefunctions, to get a propagator:
\begin{equation}\label{eq:aPhiprop}
    K_t(a,\Phi;a'\Phi')=\langle a',\Phi'|e^{-i H t}|a,\Phi\rangle = \frac{1}{2\pi |\sinh t|} \exp\left[\frac{i}{\sinh t}\left((a'\Phi + a \Phi')-(a\Phi+a'\Phi')\cosh t\right)\right].
\end{equation}
This obeys $H K_t = i\partial_t K_t$ with the initial condition and $K_t$ becomes $\delta(\Phi-\Phi')\delta(a-a')$ at $t\to 0$ (for positive or negative $t$).

\subsection{Prescription for non--essentially self-adjoint Hamiltonian}\label{ssec:nonESA}

An important subtlety coming from the restriction to the half-space $a>0$ is that $H$ is not essentially self-adjoint. In particular, this means that time evolution $e^{-i t H}$ does not unambiguously define a unitary operator on $\hilb_0$. This is not merely a technical point, since it reflects a physical phenomenon: namely, the classical theory predicts evolution to a singularity (where $a\to 0$) in finite proper time. This is of course typical for gravitational systems. Since the Hamiltonian should be imposed as a constraint we cannot simply ignore this issue, as it will enter (perhaps implicitly) in the construction of the Hilbert space. Here we give a conservative, physically-motivated prescription to deal with this.

Our attitude is that our theory (or richer theories to which we want to apply the same ideas) should not be regarded as a complete self-contained quantum theory, but as a low-energy effective approximation to a more complete theory of quantum gravity. As such, the big bang/crunch singularities are places where the description breaks down, and the more complete theory is required. This is not so obvious in the solvable pure JT gravity, but becomes more significant when we include matter (with energy density and dilton typically diverging at bang/crunch singularities).

Given this, we apply a conservative approach of maximal ignorance about the nature of singularities. This means we do not postulate any evolution beyond the singularity (or before a big bang), rather supposing that the state cannot be described in our low-energy theory after such a singularity. This means that we simply throw away any part of the state that evolves to a singularity (since it ends in a part of the unconstrained Hilbert space which we are not attempting to describe). This means that evolution $e^{-i H t}$ will be non-unitary. Very concretely, we take \eqref{eq:akevolution} to define evolution from an initial wavefunction $\psi(a,k)$ defined on $a>0$, but we extend the definition of $\psi(a,k)$ to vanish when $a<0$ (so the right-hand-side is zero when $a\cosh t+k \sinh t<0$).

This approach may not give a good description of the `low-energy' part of the Hilbert space in one circumstance: a state which evolves to a singularity  subsequently  becomes semi-classical again (with non-negligible amplitude). This can certainly happen --- for example, initial data which forms a black hole might evolve through a regime where the effective description is inapplicable, but later become a state of Hawking radiation where the low-energy theory is once again valid ---  so the physical interpretation of our prescription should be borne in mind when interpreting its results.

Note that this does \emph{not} mean that we throw away a state which may later evolve to a singularity (an approach suggested in \cite{Cotler:2023eza}), which would be far too drastic. We would not want to discard a potential state of the early universe because a black hole later forms, or because our low-energy effective theory predicts an initial big bang singularity if we run it backwards (or a final big crunch). Despite the fact that we are imposing a constraint $H$ which generates non-unitary evolution, we are nonetheless able to achieve a perfectly sensible and self-consistent description of such states.

More generally, in a semi-classical setting (where calculations use perturbation theory around classical solutions or saddle points), we expect that this prescription is implemented by a fairly standard approach of excluding configurations with strong curvature or other high-energy effects that invalidate low-energy effective field theory.

An alternative approach is to choose a different gauge for time translations which puts the singularities at infinite coordinate time. For example, a conformal gauge (choosing $N=a$ rather than our $N=1$ gauge) puts the $a\to 0$ singularities at infinite conformal time (though this introduces a new problem from the $\scri_\pm$ boundaries which occur at finite conformal time, so some hybrid gauge to combine the virtues of both may be called for).  In such a suitably chosen gauge, the Hamiltonian that generates these time translations (which is conformally related to our proper time Hamiltonian) will then be essentially self-adjoint and we will not have to face these tricky technical issues. But in fact (with appropriate definitions of the quantum Hamiltonian) this is entirely equivalent to the approach we have advocated here (for our group-average inner product in particular). We explain this gauge-independence using the BRST formalism in the companion paper \cite{Held:2025mai}. It may be technically useful to cast things in this language so that the constraint (the rescaled Hamiltonian) is a self-adjoint operator, though we do not pursue this here.

\subsection{The algebra of observables}

To organise our thinking about the Hilbert space and various nice bases of states, it will be useful to have a natural description of the algebra of observables.

We start with classical observables, namely the functions $G$ of phase space $(a,\Phi,p,k)$ such that $\{H,G\}=0$.\footnote{We could generalise to allow $\{H,G\}$ to be weakly zero, meaning that it is $H$ times something (so it vanishes when constraints are satisfied). The quantity $M$ given in \eqref{eq:Phiconst2} is an example of such an observable. But this only adds functions which are themselves weakly zero (so do not correspond to distinct physical observables) and hence does not alter the conclusions.} Because $\ham$ is quadratic in these variables, $\{\ham,\cdot\}$ maps any polynomial to another polynomial of the same degree; this means it's natural to arrange the analysis of classical observables by polynomials of fixed degree. There are no linear functions which commute with $\ham$. The first non-trivial observables are quadratic, and they are linear combinations of three generators:
\begin{equation}
    G_1 = \tfrac{1}{2}(a^2-k^2), \quad G_2 = \tfrac{1}{2}(p^2-\Phi^2),\quad G_3 = ap- \Phi k.
    \label{eq:sl2gens}
\end{equation}
These all become Hermitian operators (ordered as written) in the quantum theory.\footnote{Like the Hamiltonian, $G_2$ is not essentially self-adjoint on $a>0$, which we will have to address to define a good quantum observable.} They form an $\mathfrak{sl}(2)$ algebra:
\begin{equation}
    \{ G_1,G_2\} = G_3, \quad \{ G_1,G_3\} = 2G_1, \quad \{ G_2,G_3\} = -2G_2 \,,
\end{equation}
which remains true in the quantum theory under the usual correspondence $\{\cdot,\cdot\}\rightarrow i[\cdot,\cdot]$. 

Since the reduced phase space is only two-dimensional, we expect a single relation between these three observables. Indeed, this comes from the quadratic Casimir:
\begin{equation}\label{eq:Cas}
    C = \tfrac{1}{4}G_3^2-\tfrac{1}{2}(G_1 G_2+G_2 G_1)=\tfrac{1}{4}(H^2 + 1).
\end{equation}
We have written it this way so that this equation continues to hold in the quantum theory (with this operator ordering). Going to higher orders, the only observables are functions of $G_1,G_2,G_3$ subject to equivalence under the Casimir relation.\footnote{\textbf{Proof:} Write $q_\pm = \frac{a\pm k}{\sqrt{2}}$, $p_\pm = \frac{ p\pm \phi}{\sqrt{2}}$. In this basis, $\{q_\pm,p_\mp\}=1$ are the non-trivial Poisson brackets. We have $\ham=q_- p_+ - q_+ p_- $, so $\{\ham,\cdot\} = q_+\frac{\partial}{\partial q_+} + p_+\frac{\partial}{\partial p_+}-q_-\frac{\partial}{\partial q_-} - p_-\frac{\partial}{\partial p_-}$ counts the $\pm$ graded degree $d$: $q_+^{m_+}q_-^{m_-}p_+^{n_+}p_-^{n_-}$ picks up a factor of $d = m_++n_+-m_--n_-$. We can look for solutions to $\{\ham,f\} = \ham g$ at fixed degree $d$. For $d\neq 0$, we find that $f = \frac{g}{d} \ham$, so $f$ is trivial. So nontrivial observables must be at $d=0$. Any $d=0$ monomial can be written as a product of powers of $G_1=q_+q_-$, $G_2=p_+p_-$, and $\frac{1}{2}(G_3\pm \ham) = q_\mp p_\pm$. So functions of $G_1,G_2,G_3$ exhaust all polynomial observables. In these variables, it is also clear that the Casimir relation $(q_- q_+)(p_- p_+)=(q_- p_+)(p_- q_+)$ generates all other relations.}

\section{Wheeler-DeWitt wavefunctions}
\label{sec:WDW}

The oldest and perhaps most well-known prescription for defining the space of physical states $\Hphys$ with (first-class) constraints goes back to Dirac's foundational work \cite{Dirac:1950pj}. The physical states are simply the wavefunctions in $\hilb_0$ which are annihilated by the constraints: for us, $H|\psi\rangle=0$, which in gravity is the Wheeler-DeWitt equation \cite{DeWitt:1967yk}. We will call these states \emph{invariants}, since they are invariant under gauge transformations generated by constraints, and use the notation $|\psi\ri$ to indicate a state which solves the Wheeler-DeWitt equation. For now this is just a vector space, without the additional structure of an inner product required to turn it into a Hilbert space: we will address this later. In this section we will give a somewhat systematic classification of all such wavefunctions, choosing various natural bases and using both the $\psi(a,\phi)$ and $\psi(a,k)$ representations.


\subsection{States in the $(a,k)$ representation}\label{ssec:akStates}

It is most straightforward to solve the Wheeler-DeWitt equation in the $a,k$ basis where $H$ is a first-order differential operator \eqref{eq:Hak}. The invariant wavefunctions are those which are constant along characteristics, which are the integral curves of the vector field $a\partial_k + k\partial_a$. As commented earlier, this generates boosts in the half-Minkowski space parameterised by `space' $a$ and `time' $k$, as shown in figure \ref{fig:dSflow}. The boost naturally divides the $(a,k)$ configuration space into three sectors: the `Rindler wedge' $a>|k|$, and the upper and lower `Milne wedges' $k>a$ and $-k>a$ (with $a>0$ in each case).

In each sector, the invariant wavefunctions are independent of the boost parameter, depending only on the radial Rindler coordinate or on the Milne time. We can write this general solution in terms of three wavefunctions, each depending on a single positive variable:
\begin{equation}\label{eq:invak}
    \psi(a,k) = \begin{cases}
        \psi_+(\sqrt{k^2-a^2}) & k>a \\
          \psi_0(\sqrt{a^2-k^2}) & |k|<a \\
          \psi_-(\sqrt{k^2-a^2}) & -k>a
    \end{cases}
\end{equation}
This precisely reproduces the expectation from the direct quantisation of the classical phase space, namely the cotangent bundle of the Y-geometry in figure \ref{fig:miniPhaseSpace}.   The full physical Hilbert space decomposes as a direct sum of the three sectors:
\begin{equation}
    \Hphys=\hilb_{\mathrm{bang}}\oplus \hilb_{\mathrm{crunch}} \oplus \hilb_{\mathrm{bounce}}.
\end{equation}
This means a Wheeler-DeWitt state is determined by a triplet  $(\psi_+(\beta),\psi_0(b),\psi_-(\beta))$ of wavefunctions tkaing positive values, where
$\psi_0$ gives a wavefunction on the bounce sector (depending on the parameter $b=\sqrt{a^2-k^2}$ as in \eqref{eq:bouncesol}, which becomes the Rindler radius), while $\psi_-$ and $\psi_+$ are wavefunctions on the bang and crunch sectors respectively (depending on $\beta=\sqrt{k^2-a^2} $ as in \eqref{eq:BCsol}, which is a Milne time).


This generalises to the full two-dimensional theory. A wavefunction is a functional of $a(\theta)$, and $k(\theta)$, which defines a parameterised circle with specified intrinsic metric $a^2d\theta^2$ and extrinsic curvature. This specifies `initial data' which we expect to uniquely specify a constant curvature geometry in which our initial data surface is embedded. Since the Hamiltonian and momentum constraints are linear in $\phi$ and $p$, in this $a,k$ representation they are first order functional derivatives which generate flows in this space. This flow is the deformation of the initial data surface in the specified geometry. So, the Wheeler-DeWitt equation says that the wavefunction is a function only of the geometry of spacetime (and not of the chosen initial surface or coordinate $\theta$). In other words, physical wavefunctions are simply functions defined on the fishbone in figure \ref{fig:fishbone}. This gives one way to understand why our mini-superspace truncation is exact when we specialise to constant $a,k$. Note that this is simply the canonical version of the statement that the action is linear in the dilaton, so $\Phi$ can be integrated out as a Lagrange multiplier to give the rigid constraint $\mathcal{R}=2$ on metrics.

\subsection{States of definite $G_i$}


Recall that we identified a simple set of generators $G_i$ ($i=1,2,3$) for the algebra of observables in \ref{eq:sl2gens}, each of which commutes with the Hamiltonian constraint. We can thus choose any one of these to simultaneously diagonalise with the Hamiltonian. This means that each $G_i$ selects a natural basis of invariant states which are annihilated by $H$, and are also eigenstates of the chosen $G_i$ (labelled by the eigenvalue). Since the simultaneous eigenstates of $G_i$ and $H$ are complete in the $L^2$ inner product of the unconstrained Hilbert space, this also gives us a notion of completeness for the Wheeler-DeWitt states. Similarly a notion of (delta-function) normalisability is inherited from the unconstrained Hilbert space. These notions of completeness and normalisability will match those from the physical inner product introduced later.  We address eigenstates of each $G_i$ in turn.

\subsubsection{States of definite geometry}

First, we can consider eigenstates of $G_1 = \frac{1}{2}(a^2 -k^2)$. In the $(a,k)$ representation these are simply delta-function wavefunctions in one of the three sectors we've identified. These have the simple interpretation as eigenstates of the geometry (spacetime metric), with completely indeterminate dilaton profile. If the eigenvalue of $G_1$ is positive we can write it as $\frac{b^2}{2}$, and the relevant wavefunctions are given by \eqref{eq:invak} with $\psi_0$ a delta-function supported at $b$ (and $\psi_\pm=0$). That is,
\begin{equation}\label{eq:G1b}
    \psi_b(a,k)  = 2\delta(a^2-k^2-b^2) = \frac{1}{b}\delta(\sqrt{a^2-k^2}-b)
\end{equation}
with a normalisation constant chosen for later convenience. In the above characterisation of states, $\psi_0$ is $\frac{1}{b}$ times a delta function supported at $b$. For negative eigenvalue $G_1=-\frac{1}{2}\beta^2$ we have a similar two-dimensional space of solutions
\begin{equation}\label{eq:G1beta}
    \psi_{\beta,\pm}(a,k)  = 2\delta(a^2-k^2+\beta^2)\Theta(\pm k) = \frac{1}{\beta}\delta( \sqrt{k^2-a^2}-\beta)\Theta(\pm k) = \frac{1}{k}\delta(k\mp \sqrt{a^2+\beta^2}).
\end{equation}

We can Fourier transform these state in $k$ to give the representation as wavefunctions $\psi(a,\Phi)$. For the `bounce' states with positive eigenvalue $G_1=\frac{b^2}{2}$ we have
\begin{equation}
    \psi_b(a,\Phi) = \frac{2\cos(\Phi\sqrt{a^2-b^2})}{\sqrt{2\pi(a^2-b^2)}} \Theta(a-b),
\end{equation}
while for bang/crunch states with $G_1=-\frac{\beta^2}{2}$ we get
\begin{equation}
    \psi_{\beta,\pm}(a,\Phi) = \frac{e^{\pm i\Phi\sqrt{a^2+\beta^2}}}{\sqrt{2\pi(a^2+\beta^2)}} .
\end{equation}
We note that in the $a,\Phi$ variables we find another possible state with $H=0$ and $G_1=\frac{b^2}{2}>0$ if we simply look for solutions to these two differential equations:
\begin{equation}
    \psi(a,\Phi) = \begin{cases}\frac{\sin(\Phi\sqrt{a^2-b^2})}{\sqrt{2\pi(a^2-b^2)}} & a>b \\
        \frac{\sinh(\Phi\sqrt{b^2-a^2})}{\sqrt{2\pi(b^2-a^2)}} & a<b
    \end{cases}
\end{equation}
We exclude these states because of the rapid exponential growth at large $\Phi$ when $a<b$. This is also why we didn't see them in the $(a,k)$ representation (because an exponential is not the Fourier transform of an ordinary function or distribution). Once we discuss inner products we will have a more principled reason to discard them.

\subsubsection{States of definite dilaton invariant}\label{sssec:G2}

Next we discuss diagonalising $G_2 = \frac{1}{2}(p^2-\Phi^2)$. We can interpret this by noting that this observable corresponds to the quantity $-\frac{1}{2}M$ introduced in \eqref{eq:Phiconst} and \eqref{eq:Phiconst2} which determines the magnitude of the dilaton in an invariant way: here we write this as $M=\varphi^2$ or $M=-\tilde{\varphi}^2$ taking $\varphi>0$ or $\tilde{\varphi}>0$ (slightly different to \eqref{eq:bouncesol}, \eqref{eq:BCsol},  \eqref{eq:nAdSsol} where they take any real value).  So, $G_2$ eigenstates are states with definite dilaton profile in this sense. This means the geometry has very large fluctuations, so it remains tricky to interpret the states directly from classical intuition.

Before ploughing ahead, it is important to notice that $G_2$ is not essentially self-adjoint: the $p^2$ term is like the kinetic term for a non-relativistic particle on a half-line $a>0$, so we need to specify boundary conditions at $a=0$ to get an unambiguous self-adjoint operator with real eigenvalues, and a complete set of orthonormal eigenfunctions. One choice turns out to be most convenient and physically sensible --- in particular, producing an orthonormal basis of states under the physical inner product --- but it is not so obvious at this stage. It is most natural in the $a,k$ basis, where it amounts to discarding eigenfunctions which blow up logarithmically at the lightcones $k=\pm a$ (but allowing discontinuities there). In the $a,\Phi$ basis there is not such a natural prescription, but one possibility is to look at the boundary value $\psi(a=0,\Phi)$ and allow wavefunctions which decay as $\frac{c}{\Phi}$ as $\Phi\to \pm\infty$, but require the same coefficient in each direction (disallowing $\frac{1}{|\Phi|}$ at large $|\Phi|$, for example, since this gives singularities like $\log|k|$ in the Fourier transform).

Given the tricky boundary conditions, it is one again simplest to find valid solutions in the $(a,k)$ basis. To do this it is easiest to pass to the Rindler or Milne coordinates as in  \eqref{eq:Rindler}, where $G_2$ acts on invariant functions as $ -\frac{1}{2}(\partial_b^2 + b^{-1}\partial_b)$ in the Rindler wedge, and as $\frac{1}{2}(\partial_\beta^2 + \beta^{-1}\partial_\beta)$ in the Milne wedges. The eigenfunctions within these regions are Bessel functions. For positive eigenvalues $\frac{1}{2}\tilde{\varphi}^2$ we have solutions $J_0(\tilde{\varphi} b)$ and $Y_0(\tilde{\varphi} b)$ in the Rindler wedge, and  $I_0(\tilde{\varphi} \beta)$, $K_0(\tilde{\varphi} \beta)$ in the Milne regions. This is reversed for negative eigenvalues $-\frac{1}{2}\varphi^2$, with solutions $I_0(\varphi b)$ and $K_0(\varphi b)$ in the Rindler region and $J_0(\varphi \beta)$, $Y_0(\varphi \beta)$ in Milne regions. However, we discard all if these apart from the $J_0$ solutions, since they either grow at large $b$ or $\beta$ (for $I_0$) or have logarithmic singularities at the lightcone $b\to 0$ or $\beta \to 0$ (for $K_0$ and $Y_0$). We thus have a single eigenfunction for positive $G_2$,
\begin{equation}
    \psi_{\tilde{\varphi}}(a,k) = J_0(\tilde{\varphi}\sqrt{a^2-k^2}) \qquad (|k|<a),
\end{equation}
supported only in the bounce region, and two eigenfunctions for negative $G_2$,
\begin{equation}
    \psi_{\varphi,\pm}(a,k) = J_0(\varphi\sqrt{k^2-a^2}) \qquad (\pm k>a),
\end{equation}
supported in either the bounce or crunch sectors. 

We can now also look for eigenfunctions in the $(a,\Phi)$ representation by taking  Fourier transforms of the above $(a,k)$ wavefunctions. It is also simple to get these directly by first solving eigenvalue equations $G_2 = -\frac{1}{2}\varphi^2$  or $G_2 = +\frac{1}{2}\tilde{\varphi}^2$, and subsequently imposing the Wheeler-DeWitt equation (though the boundary conditions are not so obvious).  For positive $G_2$ our single allowed solution is
\begin{equation}\label{eq:psiG21}
    \psi_{\tilde{\varphi}}(a,\Phi) = \frac{2\sin\left(a\sqrt{\Phi^2+\tilde{\varphi}^2}\right)}{\sqrt{2\pi(\Phi^2+\tilde{\varphi}^2)}}.
\end{equation}
Another solution to the eigenvalue differential equation can be obtained by replacing $\sin$ with $\cos$, but this does not obey the boundary conditions required above (and turns out not to be linearly independent from the other eigenfunctions we include).


For negative $G_2$ our two solutions are
\begin{equation}\label{eq:psiG22}
    \psi_{\varphi,\pm}(a,\Phi) = \begin{cases} \frac{e^{-a\sqrt{\varphi^2-\Phi^2}}}{\sqrt{2\pi(\varphi^2-\Phi^2)}} & \Phi^2<\varphi^2 \\
            \pm i \frac{e^{\pm  i a\sqrt{\Phi^2-\varphi^2}}}{\sqrt{2\pi(\Phi^2-\varphi^2)}} & \Phi^2>\varphi^2.
    \end{cases}
\end{equation}
We should be careful here that there is no  contribution to $H\psi$ at the singular locus $\Phi=\pm \varphi$. To see this, note that we can obtain the solution here as the $\epsilon\to 0$ limit of an analytic function obtained by replacing $\varphi^2\to\varphi^2\mp i \epsilon$. Again, there are other reasonable-looking solutions  such as
\begin{equation}\label{eq:psiG2bad}
    \psi(a,\Phi) = \frac{2\cos(a\sqrt{\Phi^2-\varphi^2})}{\sqrt{2\pi(\Phi^2-\varphi^2)}} \Theta(\Phi-\varphi)
\end{equation}
which we are not regarding as valid eigenfunctions for $G_2$ because of the boundary conditions.

These wavefunctions are analogous to wavefunctions for fixed $M$ used in \cite{Nanda:2023wne} (following \cite{Iliesiu:2020zld}), though not quite the same. We comment on these in section \ref{sec:2D}.

\subsubsection{Eigenstates of $G_3$}\label{sssec:G3}

Next we diagonalise $G_3= ap-\Phi k$, looking for Wheeler-DeWitt states with $G_3$ eigenvalue $\lambda$. This is manifest as a symmetry on the $(a,\Phi)$ mini-superspace, which acts by  rescaling $a \mapsto \Omega a$ and $\Phi \mapsto \Omega^{-1} \Phi$. There is a three-dimensional space of such states for each $\lambda$. In the $(a,k)$ variables $G_3$ acts as a dilatation ($a \mapsto \Omega a$ and $k \mapsto \Omega k$), so the eigenstates are simple powers of the invariant quantity $a^2-k^2$, supported in one of the three wedges:
\begin{align}
    \psi_{\lambda,0}(a,k) &=\frac{1}{\sqrt{2\pi}} \left(a^2-k^2\right)^{-\frac{1}{2}+\frac{i\lambda}{2}} \Theta(a^2-k^2), \\
    \psi_{\lambda,\pm}(a,k) &=\frac{1}{\sqrt{2\pi}} \left(k^2-a^2\right)^{-\frac{1}{2}+\frac{i\lambda}{2}} \Theta(\pm k-a).
\end{align}
Fourier transforming to the $a,\Phi$ representation we have
\begin{align}\label{eq:G3aphi}
    \psi_{\lambda,0}(a,\Phi) &= \frac{1}{2\sqrt{\pi}} \Gamma \left(\tfrac{1 +i\lambda}{2}\right) \left(\tfrac{\Phi}{2a}\right)^{-\frac{i\lambda}{2} } J_{\frac{i\lambda}{2}}(a \Phi ), \\
    \psi_{\lambda,+}(a,\Phi) &=\frac{i }{4 \sqrt{\pi} }\Gamma \left(\tfrac{1+i\lambda}{2}\right)e^{-\lambda \frac{\pi}{2}} \left(\tfrac{\Phi }{2a}\right)^{-\frac{i\lambda}{2} } H^{(1)}_{\frac{i\lambda}{2} }(a \Phi), \\
    \psi_{\lambda,-}(a,\Phi) &=-\frac{i }{4 \sqrt{\pi} }\Gamma \left(\tfrac{1+i\lambda}{2}\right)e^{\lambda \frac{\pi}{2}} \left(\tfrac{\Phi }{2a}\right)^{-\frac{i\lambda}{2} } H^{(2)}_{\frac{i\lambda}{2} }(a \Phi).
\end{align}
These are correct for $\Phi>0$, but some care is required to take the correct branch to get the wavefunctions for $\Phi<0$: it is simplest to note that $\psi_{\lambda,+}(a,\Phi) = \psi_{\lambda,-}(a,-\Phi)$. This is the `Rindler basis expansion' given in \cite{Nanda:2023wne}.

\subsection{Asymptotically defined `scattering' states}\label{ssec:scatStates}

A very natural class of states in the theory are defined at a past or future asymptotic boundary $\scri_\pm$. In the unconstrained Hilbert space $\hilb_0$, these can be understood as scattering in-states and out-states with $H=0$. In the path integral, they are defined as initial and final states , boundary conditions at $t\to -\infty $ or $t\to +\infty$ with fixed $a$ and $\Phi$, where we take $a$ and $|\Phi|$ to infinity with fixed ratio $\frac{\Phi}{a} = \phi_{\mathrm{in}}$ or $\frac{\Phi}{a} = \phi_{\mathrm{out}}$  (including a boundary term in the action to obtain a finite limit). These are precisely analogous to the boundary conditions considered in \cite{Maldacena:2019cbz,Cotler_2020} and other subsequent work. The only difference for us arises from the proper interpretation of the mini-superspace truncation, where $\Phi$ is the zero-mode (spatial average of the dilaton) and $k'=0$ (so we are working on slices of constant extrinsic curvature). We comment on the change of basis to true fixed dilaton boundary conditions (and the connection to the Schwarzian theory \cite{Maldacena:2019cbz,Cotler_2020}) in section \ref{sec:2D}. Note that here we treat positive and negative $\phi_{\mathrm{in/out}}$ on an equal footing (though they have quite different interpretations when JT is interpreted as a dimensional reduction \cite{Maldacena:2019cbz})

To implement this in the canonical formalism, we can define in and out states as follows (with $\epsilon>0$ analogous to a `holographic renormalisation' cutoff):
\begin{align}
    |\phi_{\mathrm{in}}, \mathrm{in}\rangle &:= \lim_{\epsilon \to 0} \mathcal{N}_\mathrm{in}\int_0^\infty dt\, e^{-i H t} |a=\epsilon^{-1},\Phi=\epsilon^{-1}\phi_{\mathrm{in}} \rangle, \\
    |\phi_{\mathrm{out}}, \mathrm{out}\rangle &:= \lim_{\epsilon\to 0} \mathcal{N}_\mathrm{out}\int_{-\infty}^0 dt\, e^{-i H t} |a=\epsilon^{-1},\Phi=\epsilon^{-1}\phi_{\mathrm{out}}\rangle.
\end{align}
In each case we begin with an initial wavefunction $\delta(a-\epsilon^{-1})\delta(\Phi-\epsilon^{-1}\phi_{\mathrm{in/out}})$ in the $a,\Phi$ basis. We then evolve for an arbitrary time, acting with $e^{-i H t}$ where we integrate over $t$, taking forward or backward evolution ($t>0$ or $t<0$) for in- and out-states respectively. We then take the $a\to\infty$ limit, in which only the large $|t|$ tail of the integral will contribute (so the finite limit of integration at $t=0$ could be shifted to any arbitrary $t_0$ without changing the state). The normalisation factors $\mathcal{N}$ (which will have divergent phases as $\epsilon\to 0$) are chosen to obtain a finite limiting wavefunction.

Note that we can write our integrated time evolution operator formally as $\int_0^\infty dt e^{-i H t} = \frac{1}{\epsilon+i H}$, and similarly for out states. The operator $\frac{1}{\epsilon+i H}$ defines a retarded Green's function for $H$, which does not solve the Wheeler-DeWitt equation: acting with $H$ on $\frac{1}{\epsilon+i H}|\psi\rangle$ instead gives back the initial state $|\psi\rangle$ as a `source'. But we do ultimately get a state annihilated by $H$ by taking the limiting procedure, where our source is pushed to infinity.

It is easiest to carry this out in the $(a,k)$ representation, where evolution from an initial wavefunction $\psi(a,k)\frac{1}{\sqrt{2\pi}}\delta(a-\epsilon^{-1})e^{-i k \epsilon^{-1}\finout}$ is given very simply by applying a `boost' to the $a,k$ coordinates as in \eqref{eq:akevolution}:
\begin{equation}
    e^{-i H t}\psi (a,k) = \frac{1}{\sqrt{2\pi}}\delta(a\cosh t+k \sinh t-\epsilon^{-1})\exp^{-i(k\cosh t+a \sinh t)\epsilon^{-1}\finout}.
\end{equation}
The integral over $t$ picks up  a contribution from the delta-function with $(a+k)e^t \sim \epsilon^{-1}$ for large positive $t$ or $(a-k)e^{-t} \sim \epsilon^{-1}$ for large negative $t$, applying only for in-states or out-states respectively. This means that an in-state will be supported only in the region $a+k>0$, and an out-state only for $a-k>0$. That is, in states are supported on bounce and crunch sectors, while out states are supported on bounce and bang: this is not surprising, since these are the states which have asymptotic infinities in the past ($\scri_-$) and future ($\scri_+$) respectively. To do the integration it is simplest to write $a,k$ in terms of Rindler or Milne coordinates, before taking the $\epsilon\to 0$ limit. Choosing $\mathcal{N}_\mathrm{in} = \epsilon^{-1} e^{i \epsilon^{-2}\finout} = \mathcal{N}_\mathrm{out}^*$, we find in and out state wavefunctions
\begin{align}
    \psi_{\fin}(a,k) &= \frac{1}{\sqrt{2\pi}}e^{\frac{i}{2}\fin(a^2-k^2)} \Theta(a+k), \\
    \psi_{\fout}(a,k) &= \frac{1}{\sqrt{2\pi}}e^{-\frac{i}{2}\fout(a^2-k^2)} \Theta(a-k).
\end{align}
Thinking of the physical states as wavefunctions of $G_1 = \frac{a^2-k^2}{2}$, these asymptotic states are like plane waves with definite conjugate momentum. The important distinction with usual plane waves is that the negative axis $\frac{a^2-k^2}{2}<0$ splits into two branches, and the in and out states are supported on different sectors of the Hilbert space there.

While these states are rather simple in the $a,k$ representation, they do not look so nice written as wavefunctions of $a,\Phi$.

\subsection{One more basis and  a representation theory perspective}

A potentially useful perspective on the various wavefunctions presented here uses the representation theory of the $\mathfrak{sl}(2)$ algebra generated by the $G_i$ observables. The $G_i$'s are self-adjoint operators on the Hilbert space (after choosing appropriate boundary conditions for $G_2$), so the unconstrained Hilbert space becomes a unitary representation of $\mathfrak{sl}(2)$. A useful reference for the necessary representation theory is \cite{Kitaev:2017hnr}. A natural basis from this perspective comes from the decomposition into irreps, identifying states by their representation and a label for states within a given representation. The standard presentation for states within a given representation of $\mathfrak{sl}(2)$ uses a basis which diagonalises an element $L_0=\frac{1}{2}(G_1+G_2)$ (or another choice related by conjugation), which has a discrete spectrum. We will thus be led to yet another natural basis of Wheeler-DeWitt states, this time a discrete set.

We could decompose the full unconstrained Hilbert space $\hilb_0$ into irreducible representations, guided by the Casimir $C$ given in \eqref{eq:Cas}. The Casimir is related to the Hamiltonian constraint by $C=\frac{1}{4}(1+H^2)$, so the solutions to the Wheeler-DeWitt equation are all representations with $C = \frac{1}{4}$. 
The spectrum of $C$ is $[\frac{1}{4},\infty)$, so we can write its eigenvalues as $\frac{1}{4}+s^2$ with $s\geq 0$. For $s>0$, the only possible irreducible unitary representations are the principal series representations, and we see that we can identify their usual label with the energy $s=\frac{E}{2}$ (up to a possible sign). These representation are additionally labelled by $\mu\in \RR/\ZZ$, so that the spectrum of $L_0 = \frac{1}{2}(G_1+G_2)$ is $\mu + \ZZ$. We will not worry about classifying these, focusing on the physical Hilbert space of invariants (with $H=0$), which coincides with the $C=\frac{1}{4}$ subspace (if we choose boundary conditions for $G_2$ to exclude states which are annihilated by $H^2$ but not by $H$). Here we have several possible irreps appearing:  the principal series representations with $-\frac{1}{2}<\mu<\frac{1}{2}$, or discrete series representations $\mathcal{D}_{\frac{1}{2}}^\pm$ with positive or negative half-integer spectrum for $L_0$.

To determine which representations appear, it will be sufficient to work in the separate Rindler $a>|k|$ and Milne $k>a$, $k<-a$ wedges, since we have chosen boundary conditions that do not couple them. In the quadrant $a>|k|$ we use the Rindler coordinates $(b,u)$ given in \eqref{eq:Rindler} (with $b>0$) and in the `Milne' wedges similar $(\beta,v)$ coordinates, so invariant wavefunctions are given by $\psi_0(b)$ and $\psi_\pm(\beta)$. We can write the action of the $G_i$'s on $\psi_0(b)$ as follows:
\begin{equation}
G_1 = \tfrac{1}{2}b^2 ,\qquad 
G_2 = b^{-\frac{1}{2}}\left(-\tfrac{1}{2}\partial_b^2 -\frac{1}{8b^2}\right)b^{\frac{1}{2}}, \qquad  
G_3 = -ib\partial_b -\tfrac{1}{2}i,
\end{equation}
where we have written $G_2$ by conjugating with $\sqrt{b}$ to make it look more familiar. In the Milne regions, the genrators act on $\psi_\pm(\beta)$ similarly as
 \begin{equation}
G_1 = -\tfrac{1}{2}\beta^2, \qquad
G_2 = -\beta^{-\frac{1}{2}}\left(-\tfrac{1}{2}\partial_\beta^2 -\frac{1}{8\beta^2}\right)\beta^{\frac{1}{2}} ,\qquad
G_3 = -i\beta\partial_\beta -\tfrac{1}{2}i \,.
\end{equation}
%
%
Now  $G_1+G_2$ is (up to conjugation by $\sqrt{b}$ or $\sqrt{\beta}$) $\pm$ the Hamiltonian of a 3D harmonic oscillator at `angular momentum' $l=-\frac{1}{2} $ (if we look at nonzero $H$ states this generalises to $l=-\frac{1}{2} \pm i E$). Eigenstates go as a constant or $\log b$/$\log\beta$ as $b,\beta\to 0$, and our boundary condition imposes that the $\log$ term is absent. At large $b$, solutions either grow or decay like $e^{\pm \frac{b^2}{2}}$, and we allow only decay (and similarly for large $\beta$). Solutions obeying these boundary conditions are essentially the usual bound states for the harmonic oscillator, which exist only for $L_0 = \pm(n+\frac{1}{2})$ for $n=0,1,2,\ldots$, where the $+$ is for the Rindler (bounce) region and $-$ for Milne (bang/crunch) sectors. These have wavefunctions 
\begin{equation}\label{eq:L0eigenstates}
\begin{aligned}
    \psi_{n,0}(b) &= \sqrt{2} e^{-\frac{1}{2}b^2}L_n(b^2),  \\
    \psi_{n,\pm}(\beta) &= \sqrt{2} e^{-\frac{1}{2}\beta^2}L_n(\beta^2), 
\end{aligned}
\end{equation}
where $L_n$ is the Laguerre polynomial. This corresponds to a single copy of the discrete representation $\mathcal{D}_\frac{1}{2}^+$, and two copies of $\mathcal{D}_\frac{1}{2}^-$.

Now one possible condition for a sensible physical inner product is to  demand that these representations are unitary (or that the operators $G_i$ acting on invariant wavefunctions are Hermitian). This determines the inner product up to three free parameters, which are separate normalisations for these three representations.



We note that our choice of boundary condition is not completely innocuous. An alternative natural choice is to pick Neumann boundary conditions $\partial_a \psi=0$ at $a=0$. In the $b,\beta$ variables this turns out to impose that the coefficients of the $\log b$ and $\log \beta$ terms must match (which avoids singular contributions on the lightcone) and also that the coefficients of constant terms match (since if they so not, we get $\partial_a \psi|_{a=0}\propto \delta(k)$). The allowed solutions are then
\begin{equation}
     \psi_m(a,k) = \begin{cases}
         \frac{\sqrt{2}}{\pi}\Gamma\left(\tfrac{1}{2}-m\right)e^{-\frac{1}{2}(a^2-k^2)} U\left(\tfrac{1}{2}-m,1,a^2-k^2\right)  \quad a>|k|\\
         \frac{\sqrt{2}}{\pi}\Gamma\left(\tfrac{1}{2}+m\right)e^{-\frac{1}{2}(k^2-a^2)} U\left(\tfrac{1}{2}+m,1,k^2-a^2\right)  \quad a<|k|
     \end{cases}
\end{equation}
for integer $m$. From the eigenvalues and Casimir, we can identify this as a single copy of the principal series representation $\mathcal{C}^0_\frac{1}{4}$ (in the notation of \cite{Kitaev:2017hnr}). And indeed, we can check that these functions fall into that representation (i.e., the operators $G_1-G_2\pm i G_3$ act as raising and lowering operators). When it comes to imposing an inner product, this is more restrictive than the above because we only have a single irreducible representation, and hence a unique inner product up to overall normalisation. But it turns out to disagree with the  group averaging proposal that we introduce in the next section! This is one reason to prefer the definition of $G_2$ that we have adopted.

\section{Inner products from group averaging}
\label{sec:IP&GA}
We now have a fairly complete systematic understanding of the space of Wheeler-DeWitt wavefunctions  of the theory. But this does not yet constitute a Hilbert space (and allow us to calculate physical observables) because we are missing a crucial ingredient: an inner product. In this section we define an inner product using the `group averaging' prescription, a natural way to define a manifestly positive-definite inner product.

\subsection{Co-invariant states}\label{ssec:coinv}

Before discussing the inner product on invariant Wheeler-DeWitt states $\hinv$, it will be useful to first introduce an alternative representation of the physical Hilbert space (later explaining the equivalence between the two possibilities). The idea is that we do not any longer impose a constraint $H|\psi\rangle$ directly on the wavefunctions (giving invariant states $|\psi\ri$),  instead allowing any wavefunction but imposing an equivalence relation identifying different wavefunctions which differ by a gauge transformation:
\begin{equation}\label{eq:coinvequiv}
    |\psi\rangle \sim |\psi\rangle + H|\phi\rangle.
\end{equation}
With this prescription, physical `co-invariant' states (which we write as $|\psi\rc$) are equivalence classes (or cosets)
\begin{equation}
    |\psi\rc := \{|\psi\rangle + H|\phi\rangle\} \in \hco.
\end{equation}
We can also consider identifying states under finite gauge transformations, here $|\psi\rangle\sim e^{-iHt}|\psi\rangle$. This is equivalent to the above if the constraints are self-adjoint, though for us will be slightly problematic because of the boundary $a=0$. States in the image of $H$ can be thought of as `null' or `pure gauge', equivalent to the zero wavefunction. We will sometimes slightly abuse notation and write $|\psi\rc$ for a particular state in the unconstrainted Hilbert space $\hilb_0$, where we want to emphasise that it should be regarded as a  particular representative state of a co-invariant coset.

The name `co-invariant' comes about because $\hco$ is dual (in the sense of linear algebra) to the space $\hinv$ of invariants. That is, there is a natural pairing $\li \psi'|\psi\rc$ between an invariant Wheeler-DeWitt state $|\psi'\ri$ and a  co-invariant equivalence class $|\psi\rc$. This is because the original inner product $\langle \psi'|\psi\rangle$ in the unconstrained Hilbert space does not depend on the choice of representative $|\psi\rangle$: adding a null state $H|\phi\rangle$ does not change the pairing because $\langle \psi'|H|\phi\rangle=0$, by taking (Hermitian) $H$ to act to the left. (In a more technically careful treatment of the formalism, this duality is used to \emph{define} $\hinv$ as the dual space of $\hco$, since the invariant states are distributional.)

Just like for the invariant states, it will be very easy for us to understand the space of co-invariant states using the $(a,k)$ basis, and even more so in the Rindler coordinates $(b,u)$ or Milne coordinates $(\beta,v_\pm)$. In that representation the Hamiltonian constraint acts (in the Rindler wedge) simply as $i\partial_u$, generating translations in $u$. Any wavefunction which can be written as a partial derivative with respect to $u$ is therefore null. This is any state for which the integral $\int_{-\infty}^\infty du\, \psi(b,u)=0$ vanishes for each $b$ so that $\phi(b,u) = -i\int_{-\infty}^u du\, \psi(b,u)$ gives a good state (requiring $|\phi\rangle$ in \eqref{eq:coinvequiv} to decay  sufficiently rapidly at infinity).\footnote{Mathematically, we can start by defining the co-invariants starting with a dense subspace of `nice' test function wavefunctions, perhaps smooth wavefunctions of compact support in the $(a,\Phi)$ half-plane $a>0$. We ultimately give a final definition of the co-invariant Hilbert space after defining the group-averaging inner product, and taking the completion with respect to this norm.} Therefore, the space of cosets $\hco$ is equivalent to the space of functions $\int_{-\infty}^\infty du\, \psi(b,u)$ of $b$ (the only data left invariant by adding an arbitrary pure-gauge state $H|\phi\rangle$). It is tempting to identify this function with the wavefunction $\psi_0(b)$ in \eqref{eq:invak}, and we will see roughly this prescription (up to an important normalisation) from a more principled approach below.

Similar considerations apply also in the Milne wedges, if we restrict our states to have compact support. In particular we require all the wavefunction in the present discussion to vanish in a neighbourhood of the potentially problematic $a=0$ boundary, so that the integrals $\int_0^\infty dv_\pm \psi(\beta,v_\pm)$ as a function of $\beta$ (along with the above function of $b$) give an invariant representation of the co-invariant states.

A useful aspect of the co-invariant representation is that it gives us a convenient way to define states in the bulk, in some gauge-fixed way. This corresponds to states defined in the path integral by boundary conditions at finite time. For example, we can define geodesic fixed-length states $|b\rangle$ by delta-function wavefunctions
\begin{equation}\label{eq:geostate}
    \langle a,k|b\rangle = \delta(a-b)\delta(k).
\end{equation}
A superposition of such states with different values of $b$ can be thought of as a wavefunction with a $k=0$ gauge-fixing.
While these are not physical wavefunctions of the constrained theory  \emph{a priori}, we can immediately think of them as physical states by associating them with the coset $|b\rc$ (much as a classical configuration of a gauge field can be specified simply by choosing one particular representation in a specific gauge). Furthermore, such a state is very naturally defined in the path integral by a geodesic boundary, where we impose boundary conditions fixing $k=0$ and proper length $a=k$. 

We can also take a superposition of geodesic states, with wavefunction $\psi(a,k) = \delta(k)\psi_\perp(a)$ for some `transverse' wavefunction $\psi_\perp(a)$. The co-invariant states of this form precisely correspond to the bounce sector of $\hco$: any (rapidly decaying) wavefunction $\psi(a,k)$ with support contained in the region $a>|k|$ is equivalent under \eqref{eq:coinvequiv} to a \emph{unique} state of this form (with $\psi_\perp$ given by the integral described above). This is a quantum analogue of the statement that $k=0$ is a good gauge choice fixing the Hamiltonian constraint in the bounce sector.

More generally, we can attempt to define gauge-fixed co-invariant states with any gauge condition $F=0$ by wavefunctions $\psi = \delta(F)\psi_\perp(q_\perp)$, where $q_\perp$ are some variables transverse to the gauge-fixing function $F$ (and more generally still, it may be convenient to replace $\delta(F)$ with some other less singular function such as a narrow Gaussian of $F$). This is a `good gauge' if every co-invariant coset of interest contains exactly one representative of the chosen form. If there is more than one representative we have identifications between different $\psi_\perp(q_\perp)$, which can be thought of as residual gauge transformations. We might also think of $F$ as a `clock field', and working out how shifts of $F$ change $\psi_\perp(q_\perp)$ while remaining in the same co-invariant coset gives us  a relational notion of time-evolution. These ideas provide much of our motivation  for using co-invariant states, though we do not explore them much further in this paper.



\subsection{Inner products and rigging maps}

Now to define an inner product on the space of co-invariant wavefunctions, we can try to write it as the matrix elements of some operator $\eta$ in the unconstrained Hilbert space,
\begin{equation}
    \lc \psi'|\psi\rc := \langle \psi'|\eta|\psi\rangle,
\end{equation}
where on the right hand side $|\psi\rangle,|\psi'\rangle\in\hilb_0$ are elements of the co-invariant equivalence classes $|\psi\rc$, $|\psi'\rc$ respectively. For this to be an inner product, $\eta$ should be Hermitian and positive semi-definite. But more importantly, to be well-defined it must also give the same result for any choice of $|\psi\rangle,|\psi'\rangle$, so it must give zero for null states. This means $\eta H = H\eta=0$. Such an $\eta$ is known as a `rigging map'.

Note that another statement of the same condition is that $\eta$ is a map from $\hco$ to $\hinv$: it takes any state in $\hilb_0$ and produces a solution of the Wheeler-DeWitt equation ($H\eta=0$), and annihilates null states ($\eta H=0$) so it is well-defined on $\hco$. Under the further condition that $\eta$ is non-degenerate (regarded as a  map from $\hco$ to $\hinv$), it becomes a linear isomorphism between $\hco$ and $\hinv$. Thus, our two different ways of defining physical states are rendered equivalent.

We can then use this isomorphism to define an inner product on Wheeler-DeWitt states $\hinv$. To compute the inner product $\li \psi'|\psi\ri$ between invariant states, we first identify the co-invariant state $|\psi\rc$  which is mapped to $|\psi\ri$ by the rigging map $\eta$. Concretely, this means finding a specific representative wavefunction $|\psi\rangle$ which $\eta$ takes to the state of interest (for which there are many possible choices). To reiterate the potentially confusing notation here: $|\psi \rangle$ is some state in the unconstrained Hilbert space (a wavefunction of $a,\Phi$), $|\psi\rc$ is the co-invariant equivalence class containing  $|\psi \rangle$, and $|\psi\ri=\eta |\psi\rangle$ is the solution of the constraints obtained by acting with the rigging map $\eta$. Once we have found a state $|\psi\rangle$ corresponding to $|\psi\ri$, the canonical pairing $\li \psi' |\psi\rc$ defines an inner product on the original invariants:
\begin{equation}\label{eq:invIP}
    \li \psi'|\psi\ri := \li \psi'|\psi\rc, \qquad \text{where }\eta |\psi\rc= |\psi\ri.
\end{equation}
We note that this is slightly abstract and non-constructive (since we have not specified a specific `inverse' to $\eta$ which computes $|\psi\rc$). For us the $(a,k)$ representation makes things very simple so this will not be much of a problem, but more generally it may not be so obvious how to  do this.

This motivates a more direct expression in terms of invariants where $\li \psi'|\psi\ri$ is given directly in  terms of matrix elements of some operator $\kappa$ between invariant wavefunctions. In the same way that $\eta$ is a map from co-invariants to invariants, $\kappa$ should be an inverse map from invariants to co-invariants. But because a co-invariant $|\psi\rc$ can be represented by many representative states $|\psi\rangle$, there is a large amount of freedom in $\kappa$. The requirement is that $\kappa$ is  a \emph{generalised} inverse to $\eta$:
\begin{equation}
    \begin{gathered}
        |\psi\rangle \xrightleftharpoons[ \quad\kappa\quad]{\eta} |\psi\ri, \qquad \eta \kappa \eta = \eta  
    \end{gathered}
\end{equation}
One way to read this equation is that $\kappa$ is a right inverse to $\eta$ ($\eta\kappa= 1$) when acting on invariants (the image of $\eta$). Another is that the matrix elements of $\kappa$ between the invariant states $\eta|\psi\rangle$ and $\eta|\psi'\rangle$ reproduces the desired co-invariant inner product $\lc \psi'|\psi\rc=\langle \psi'|\eta|\psi\rangle$. In general it is tricky to explicitly find such a $\kappa$: we will discuss this in the next section, relating it to the problem of finding a good gauge-fixing condition.

\subsection{Group averaging}

It remains to choose a rigging map $\eta$. If we had an algebra of constraints which generated a compact gauge group, there would be an obvious candidate for a Hermitian map $\eta$ taking a general state onto a gauge-invariant state: the projector onto the gauge-invariant subspace. For our non-compact gauge group $\RR$ (time translations generated by $H$, which has continuous spectrum) this projector does not strictly exist, but there is a morally similar operator which sends a state to its component that solves the constraint: $\delta(H)$. To generalise this to a large class of possible gauge transformations, note that one way to write the projector onto singlets for a compact gauge group is the integral over the group. This generalises to many cases of non-compact gauge groups with a `group average'. For a single constraint $H$, this is simply
\begin{equation}
    \eta = 2\pi\delta(H) = \int_{-\infty}^\infty  dt e^{-i H t}\,.
\end{equation}

This is a good candidate for an inner product because it is manifestly Hermitian, positive semi-definite (because $\delta$ is a positive distribution), and compatible with the constraints  satisfying $H\eta=\eta H=0$. The group average inner product between co-invariant states (such as the states with fixed $a,k$ or $a,\Phi$ above) also has natural interpretation in the path integral formalism of gravity, described  in the next subsection.

Let's see how the group averaging works explicitly for JT by applying to the simple example of our geodesic states $|b\rc$ given in \eqref{eq:geostate}. We can compute the $(a,k)$ wavefunction of $\eta|b\rc$ as
\begin{equation}\label{eq:etab}
\begin{aligned} 
    \langle a,k|\eta |b\rc &= \int_{-\infty}^\infty dt \, \delta(a\cosh t+k \sinh t-b)\delta(k\cosh t +a \sinh t)\\
    &= \frac{1}{b}\delta(\sqrt{a^2-k^2}-b),
\end{aligned}
\end{equation}
where the second $\delta$-function gives a single contribution to the integral from $t=-\tanh\frac{k}{a}$ in the `bounce' region $k^2<a^2$ (and we get zero otherwise). By taking the overlap with another geodesic wavefunction $|b'\rc$ (which amounts to simply evaluating at $k=0,a=b'$), we immediately get the group-averaged inner product of geodesic states:
\begin{equation}
    \lc b'|b\rc = \frac{1}{b}\delta(b-b').
\end{equation}

Note that our result \eqref{eq:etab} for the invariant image $|b\ri = \eta|b\rc$ of a geodesic state is precisely the wavefunction $\psi_b$ given in \eqref{eq:G1b}. Thus, the definition \eqref{eq:invIP} of the group averaging inner product on invariants also gives us
\begin{equation}
    \li b'|b\ri = \frac{1}{b}\delta(b-b').
    \label{eq:G1norm}
\end{equation}
Since these states span the `bounce' sector, we can write the inner product of two invariant states $|\psi\ri,|\psi'\ri$ restricted to this sector in terms of their $\psi_0(b)$ wavefunction as defined in \eqref{eq:invak} (so the invariant state wavefunction in the $a,k$ sector is $\psi_0\sqrt{a^2-k^2})$ supported in the region $a>|k|$). We have
\begin{equation}\label{eq:GAbounce}
    \li \tilde{\psi}|\psi\ri = \int_0^\infty b db \tilde{\psi}_0^*(b) \psi_0(b).
\end{equation}
This $\int b db$ integral measure may be familiar from the AdS JT path integral, where it appears when gluing surfaces with geodesic boundaries of length $b$: this is no coincidence, as the gluing measure indeed can be interpreted as this inner product in a Hilbert space of closed universes (see also \ref{ssec:kIP}).

In the other (bang/crunch) sectors, subtleties arise because of the $a=0$ boundary. But with the prescription discussed in section \ref{ssec:nonESA}, we can straightforwardly apply group averaging and get physically sensible results (which do not depend in detail on the physics associated with the singularities). For co-invariants $|a,k\rc$ represented by states of definite $a,k$ (which form a basis of all of $\hilb_0$), we have
\begin{equation}
    \lc a,k|a',k'\rc = \begin{cases}
        2\delta\left((a^2-k^2) -(a'^2-k'^2)\right) & a^2-k^2>0 \text{ or }\sgn k=\sgn k', \\
        0 &\text{otherwise},
    \end{cases}
\end{equation}
where the conditions exludes a contribution when $(a,k)$ and $(a,k')$ lie in opposite Milne wedges.  We can also read this equation as giving the wavefunction (depending on $a,k$) of the invariant state  $|a',k'\ri = \eta|a',k'\rc$ obtained by group averaging of a fixed $a=a',k=k'$ state (or a more general state by taking superpositions of these). For $a'^2-k'^2=b^2>0$, this is the state $|b\ri$ as given above and in \eqref{eq:G1b}. For $a'^2-k'^2=-\beta^2<0$ this is one of the states $|\beta,\pm\ri$ with wavefunction given in \eqref{eq:G1beta}. Any wavefunction $\delta(a-a_0)\delta(k-k_0)$ with $k_0^2-a_0^2=\beta^2$ and the correct sign of $k_0$ (or a superposition of such states) is a representative of the corresponding co-invariant state $|\beta,\pm\rc$.

The corresponding inner product on invariants can be written in terms of wavefunctions $\psi_0(b)$, $\psi_+(\beta)$, $\psi_-(\beta)$ as in \eqref{eq:invak}. The norm of a state is
\begin{equation}\label{eq:GAinv}
    \li \psi|\psi\ri = \int b db |\psi_0(b)|^2+\int \beta d\beta |\psi_+(\beta)|^2 + \int \beta d\beta |\psi_-(\beta)|^2 \,.
\end{equation}
This is a sensible physical inner product, which is positive-definite on the space of solutions to the Wheeler-DeWitt equation, and imposes no further constraints on the space of allowed wavefunctions.

\subsection{Group averaging and the path integral}

For our theory of pure JT, paritcularly in the $(a,k)$ basis, the group averaging calculation is deceptively simple and does not generalise readily in the way we have presented it. With this in mind, we briefly comment on the evaluation of the rigging map $\eta$ in the path integral.

The matrix elements of $\eta$ have a very simple path integral interpretation. Since $e^{-i H t}$ is a time-evolution by proper time $t$, it is evaluated by a path integral on a spacetime interval with metric specified by the single `modulus' $t$ (which we can think of as the lapse $N$ if we fix the coordinate length of the interval). This is the only gauge-invariant data that determines a one-dimensional metric on the interval (analogous to the  moduli space parameters of the string wordsheet). By integrating over $t$, we are simply performing a path integral over all one-dimensional metrics on an interval, modulo diffeormorphisms. This becomes a general path-integral prescription for generalising $\eta$ in higher dimensions.

Note, however, that it is essential that we integrate over all real $t$: negative $t$ corresponds to `backwards' time evolution. If we do not do this, we get a `propagator' $\frac{1}{\epsilon + i H}$, rather than $\delta(H)$, which does not project onto solutions of the constraints.  We used such a propagator in section \ref{ssec:scatStates} to define the scattering states, but pushed the initial (or final) state to infinity so that could still get an invariant state: this can be thought of as choosing spacetime to be a semi-infinite line rather than a finite interval.

Also, note that this prescription works in Lorentzian signature. Since our Hamiltonian is not bounded from above or below, it is hard to make sense of a Euclidean evolution operator $e^{-H\tau}$. Even if we could modify $H$ to have  spectrum bounded from below (e.g., by a `contour rotation'  as commonly used to circumvent the conformal factor problem), a Euclidean path integral does not appear to allow the analogous `negative lapse' states, so we would again end up with a propagator (like a cylinder diagram in string theory). See further comments in \cite{Casali:2021ewu}.

In a semi-classical limit this path integral is dominated by classical solutions obeying all equations of motion, including the constraints. For fixed $t$ the path integral computes the matrix elements of $e^{-i H t}$, which is dominated by solutions which obey classical equations of motion but not the constraints, giving $e^{iS}$ for on-shell action $S$. The integral over metrics (i.e., over proper time $t$) is then dominated by paths for which $\frac{\partial S}{\partial t}$ (at fixed endpoints) vanishes. But this quantity is minus the energy of the trajectory, so the stationary phase points correspond to solutions also satisfying the constraints.

To illustrate this concretely, let's calculate the group-average inner product in the $a,\Phi$ basis, $\lc a_2,\Phi_2|a_1,\Phi_1\rc = \langle a2,\Phi2|\eta|a_1,\Phi_1\rangle$, using the semi-classical limit (which means taking large $a,\Phi$). For a fixed total proper time $T$, there is a unique solution to the equations of motion for $a,\Phi$  with these boundary conditions,
\begin{equation}
    a(t) = \frac{a_1 \sinh(T-t) +a_2\sinh(t)}{\sinh T}\, , \quad \Phi(t) = \frac{\Phi_1 \sinh(T-t) +\Phi_2\sinh(t)}{\sinh T},
\end{equation}
with on-shell action
\begin{equation}
    S = -\int_0^T dt\, (\dot\Phi\dot a+a\Phi) = \frac{(a_1\Phi_2+a_2\Phi_1)-(a_1\Phi_1+a_2\Phi_2)\cosh T}{\sinh T}.
\end{equation}
This already appeared as the exponent in the exact propagator \eqref{eq:aPhiprop} (and the prefactor is the one-loop determinant). Now $\frac{\partial S}{\partial T}$ is minus the energy $E = -\dot\Phi\dot a+a\Phi = \frac{(a_1\Phi_2+a_2\Phi_1)\cosh T-(a_1\Phi_1+a_2\Phi_2)}{\sinh^2 T}$. Now there are either two solutions to $E=0$ (one being the negative of the other) or none. In the regime where there are two solutions, they both contribute in the stationary phase approximation to the $T$ integral, with action
\begin{equation}
    S = \pm\sqrt{(a_1^2-a_2^2)(\Phi_1^2-\Phi_2^2)}.
\end{equation}
Including the one-loop fluctuation, we find the matrix of inner products
\begin{equation}
    \lc a_2,\Phi_2|a_1,\Phi_1\rc \sim  \frac{1}{\sqrt{2\pi i}} ((a_1^2-a_2^2)(\Phi_1^2-\Phi_2^2))^{-1/4} e^{-i\sqrt{(a_1^2-a_2^2)(\Phi_1^2-\Phi_2^2)}} + \text{c.c}.
\end{equation}
This could be used, for example, to compute inner products of coherent states with Gaussian wavepackets with small fluctuations in $a,\Phi$ and also their conjugate momenta, which correspond to classical spacetimes.

%
If there are no real classical solutions with zero energy (for example $a_1>a_2$ and $\Phi_2>\Phi_1>0$), the integral over $t$ can be computed by a saddle-point at a complex value of $t$, and gives an exponentially small contribution \cite{Marolf:1996gb}. This can be understood as a saddle-point with a complex metric, though it is used to calculate  an integral originally defined over ordinary real Lorentzian spacetimes.

\subsection{Inner  products of Wheeler-DeWitt states}

Now we apply our group averaging to the various states appearing in section \ref{sec:WDW}. We'll use these results as a point of comparison when we discuss alternate definitions of the inner product on $\hinv$ in the next section.

\subsubsection{Overlap of asymptotic scattering states}

First, let us look at the overlaps of the most physically motivated states, namely the asymptotic in states (defined at $\scri_-$) and out states (defined at $\scri_+$) described in section \ref{ssec:scatStates}.

From inspection of the wavefunctions of these states in the $a,k$ basis, we can write them in terms of the phsyical wavefunctions $\psi_0$, $\psi_\pm$ as in \eqref{eq:invak}.
\begin{equation}
    |\fin\ri \rightarrow \begin{cases} \psi_0(b)= \frac{1}{\sqrt{2\pi}}e^{\frac{i}{2} \fin b^2} \\
    \psi_+(\beta)= \frac{1}{\sqrt{2\pi}}e^{-\frac{i}{2} \fin \beta^2} \\
    \psi_-(\beta)= 0
    \end{cases} \qquad |\fout\ri \rightarrow \begin{cases} \psi_0(b)= \frac{1}{\sqrt{2\pi}}e^{-\frac{i}{2} \fout b^2} \\
    \psi_+(\beta)= 0 \\
    \psi_-(\beta)= \frac{1}{\sqrt{2\pi}}e^{\frac{i}{2} \fout \beta^2}
    \end{cases}
\end{equation}
We can then take the overlap of two in states using \eqref{eq:GAinv},
\begin{equation}
    \begin{split}
     \ibraket{\fin'}{\fin}&= \frac{1}{2\pi}\int_0^{\infty}b db\,\,e^{\frac{i}{2}(\fin-\fin')b^2} +\frac{1}{2\pi}\int_0^{\infty}\beta d\beta\,\,e^{-\frac{i}{2}(\fin-\fin')\beta^2} \\
     &=\int_{-\infty}^{\infty}\frac{dg_1}{2\pi}\,e^{ i (\fin-\fin')g_1} =\delta(\fin-\fin'),
     \end{split}
\end{equation}
where the second line combines the two integrals into a single integral over eigenvalues $g_3$ of $G_3=\frac{a^2-k^2}{2}$. We therefore see that the in-states form an orthonormal basis of the bounce and crunch sectors (though they do not touch the big bang sector). Similarly, the out-states are orthonormal with $\li\fout|\fout'\ri = \delta(\fout-\fout')$.

For the overlap between in and out states, they interact only in then bounce sector where geometries contain  both past and future infinities $\scri_-$ and $\scri_+$. We find
\begin{equation}\label{eq:S-matrix}
    \begin{aligned}
        \li \fout|\fin\ri = \frac{1}{2\pi}\int_0^\infty b db e^{\frac{i}{2}(\fin+\fout)b^2} = \frac{i}{2\pi}\frac{1}{\fin+\fout+i\epsilon}.
    \end{aligned}
\end{equation}
This reproduces the result of \cite{Cotler:2019dcj,Cotler:2023eza} from the path integral. In particular, the $i\epsilon$ here follows from our definition of the states in section \ref{ssec:scatStates} (prescribing states in the distant past or future), justifying a prescription which was previously somewhat ad hoc.

\subsubsection{Inner products of $G_i$ eigenstates}
We can also compute the inner products of the bases of $G_i$ eigenstates above, finding in each case that they are orthonormal. This shows that the operators $G_i$ are Hermitian operators on the physical Hilbert space, equipped with the group-average inner product.

We have already computed the overlap of $G_1$ eigenstates in \ref{eq:G1norm} and the following discussion. The $G_3$ eigenstates are straightforward, with
\begin{equation}\label{eq:G3GA}
\begin{aligned}
     \li \lambda',0|\lambda,0\ri &= \frac{1}{2\pi}\int_\infty  b db \left(b^2\right)^{-1+\frac{i}{2}(\lambda-\lambda')} \\
     &= \frac{1}{2\pi}\int_{-\infty}^\infty dx\, e^{i(\lambda-\lambda')x} \qquad \left(x=\log b\right)\\
     &= \delta(\lambda-\lambda'),
\end{aligned}
\end{equation}
And similar calculations for $\li \lambda',\pm|\lambda,\pm\ri$. These states are  manifestly orthogonal ($\li \lambda',0|\lambda,\pm\ri=\li \lambda',-|\lambda,+\ri=0$) since they have support in different sectors. For the $G_2$ eigenstates, we use the `closure identity' for Bessel functions,
\begin{equation}
    \int_0^\infty \beta d\beta J_0(\varphi \beta)J_0(\varphi' \beta) = \frac{\delta(\varphi-\varphi')}{\varphi},
\end{equation}
which tells us $\li \varphi',+|\varphi,+\ri = \varphi^{-1}\delta(\varphi-\varphi')$ and similarly for other sectors. Finally, the $L_0=\frac{1}{2}(G_1+G_2)$ eigenstates $|n,\pm\ri$ and $|n,0\ri$ as defined in \eqref{eq:L0eigenstates} are orthonormal, $\li n',\pm|n,\pm\ri = \delta_{nn'}$.

\section{Klein-Gordon inner product and generalisations}
\label{sec:KG+}

The group averaging procedure has given us a very satisfactory inner product, leading to a physically sensible Hilbert space in line with expectations from a direct quantisation of the classical phase space. However, the definition of the inner product on Wheeler-DeWitt states was indirect, requiring us to find an inverse to the group-average rigging map $\eta$. While this was straightforward for our simple model (making use of the $(a,k)$ representation with first-order constraints), it may not be so easy for more interesting and complicated theories (where we likely have access only to some semi-classical expansion). So, it is useful to have a more direct and constructive approach to defining and computing the inner product of invariant wavefunctions. This section discusses constructions which aim to achieve that goal.

\subsection{Klein-Gordon inner product and gauge-fixing}

There is an example of an `inner product' with a simple direct expression in terms of  Wheeler-DeWitt wavefunctions, and it is probably much more well-known than the group-averaging used above: the Klein-Gordon inner product \cite{DeWitt:1967yk,Wald:1993kj}. Focusing on mini-superspace models, the Hamiltonian constraint takes the form $H=-\frac{1}{2}\nabla^2 + V$, where $\nabla^2$ is the Laplacian on mini-superspace with some (typically Lorentzian) metric (here $ds^2 = -2d\Phi da$), and $V$ is a potential function on superspace (here $V=a\Phi$). The Wheeler-DeWitt equation is like a Klein-Gordon equation with potential giving a non-constant mass $V=\frac{m^2}{2}$. By analogy to construction of single-particle states in free QFT, the Klein-Gordon form is defined for some choice of (usually spacelike) hypersurface $\Sigma$ in mini-superspace, and we can write it as
\begin{equation}\label{eq:KG}
    \li \psi_2|\psi_1\ri_{\mathrm{KG}(\Sigma)} = \frac{1}{2i} \int_\Sigma \left( \psi_2^* \nabla_n \psi_1- (\nabla_n \psi_2^*) \psi_1 \right).
\end{equation}
The integral over $\Sigma$ uses the measure from the induced metric, and $\nabla_n$ is the normal derivative (we must also specify which direction $n$ points in, which changes the inner product by a sign). This is invariant under deformations of the surface $\Sigma$ (at least for compactly supported deformations) using the Wheeler-DeWitt equation satisfied by $\psi_{1,2}$.

However, this is problematic as a physical inner product for reasons which are also very well-known, most importantly that it is not positive-definite. This means that it certainly does not coincide with the group averaging inner product. Nonetheless, its failures and its relation to the group-averaging inner product have a simple interpretation. In particular, this interpretation leads to predictable circumstances (illustrated explicitly below) where the Klein-Gordon inner product does agree with the physical group-averaging inner product.


The key idea is to interpret the Klein-Gordon inner product as arising from a gauge-fixing procedure. Its shortcomings arise because the corresponding gauge choice is not always good. In a mini-superspace model the relevant gauge symmetry we need to fix is the residual time-translation generated by $H$ (which  is what remains of diffeomorphisms after fixing to constant lapse $N=1$, for example), or the choice of an origin $t=0$ for the time coordinate. In general, we might consider a gauge choice $F=0$, where $F$ could be any function of phase space variables (or a Hermitian operator with continuous spectrum including $0$) which we specify to vanish at $t=0$.  To implement such a gauge-fixing, we can insert an operator $\kappa_F$ defined (at least formally) by the time derivative of a step function $\Theta(F)$,
\begin{equation}\label{eq:kappaF}
    \kappa_F = i [H,\Theta(F)] \sim \frac{\partial F}{\partial t} \delta(F).
\end{equation}
The second expression $\frac{\partial F}{\partial t} \delta(F)$ is a useful heuristic for this operator: it gives the classical limit, or the functional one would insert in the path integral (though it is not precisely correct due to operator ordering problems when $[H,F]$ does not commute with $F$). In the path integral, $\delta(F)$ is the gauge-fixing insertion and $\frac{\partial F}{\partial t}$ is the Fadeev-Popov determinant (here just one-dimensional) of the gauge-variation of the gauge-fixing conditions. Formally, under small variations of  $F$ this expression is unchanged between invariant wavefunctions (annihilated by $H$).\footnote{One might think that \eqref{eq:kappaF} itself already vanishes between states satisfying $H=0$ since $H$ can act either to the left or right in the two terms of the commutator. This is too fast because these wavefunctions are not really in the Hilbert space since they do not decay (to make $H$ act  to the left we integrate by parts, but this can give boundary terms which do not vanish). However, changes in $F$ can give an operator which is compactly supported in mini-superspace, in which case $\delta\kappa_F$ does vanish between invariants. For our JT model, this can also fail from boundary terms at $a=0$.} This is the usual statement that the Fadeev-Popov determinant gives an invariant measure. In the Hamiltonian BFV-BRST formalism, $\kappa_F$ arises from the exponential of a BRST-exact operator ($[Q_B,\Psi]_+$ for a `gauge-fixing fermion' $\Psi$), which we will explain in \cite{Held:2025mai}. However, this may not always produce the correct result as we will explain momentarily.

For the Klein-Gordon \eqref{eq:KG}, we choose $F$ to be a coordinate on mini-superspace which vanishes on $\Sigma$ (and increases in the direction of the normal $n$). See \cite{witten2023notecanonicalformalismgravity} for more details in the path integral formalism, and \cite{Held:2025mai} for a detailed discussion in the language of Hamiltonian BRST.

If we consider \eqref{eq:kappaF} in the path integral, for a given trajectory there will be a  positive contribution each time the surface $F=0$ is crossed from negative values to positive, and negative each time $F=0$ is crossed in the reverse direction. The correct positive gauge-invariant measure would require a factor of $|\partial_t F|$, not just $\partial_t F$ as appears in the Klein-Gordon expression (or would come from a ghost determinant).\footnote{One might try to explicitly include an absolute value $|\partial_t F|$ to get a good gauge-fixed measure (e.g., \cite{Marolf:1996gb}). We expect this to be problematic in the quantum theory, since the typical trajectory in the path integral is highly erratic and crosses a given surface infinitely many times (so the result will be highly UV sensitive). But this `fixing by hand' should not cause any issues in perturbation theory.} So, $\kappa_F$ gives a good gauge-fixing operator for trajectories which begin at negative $F$ and end at positive $F$ (crossing the gauge-fixing slice once `net', where we count crossings with signs). Trajectories which cross in the reverse direction of decreasing $F$ will count negatively, so this will give us states with negative norm. Finally, there may be trajectories which do not cross the $F=0$ slice at all, and these will not by counted by $\kappa_F$.\footnote{If $F$ itself is not globally defined, for example a periodic coordinate, we could also have cases where a single trajectory is counted multiple times.}

These considerations account for the failures of the Klein-Gordon inner product, but also for its successes where it does provide a useful representation of the physical (group-average) inner product. Namely, for calculations where all trajectories cross the gauge slice $\Sigma$ in the same direction, the Klein-Gordon inner product is expected to agree with the group average inner product (at least to all orders in perturbation theory where the classical gauge-fixing intuition applies). But the Klein-Gordon expression may be useful since it is often easier to calculate. We illustrate this with some example calculations.

\subsection{Calculations of Klein-Gordon norm}

\begin{figure}
    \centering
    \includegraphics[width=0.4\linewidth]{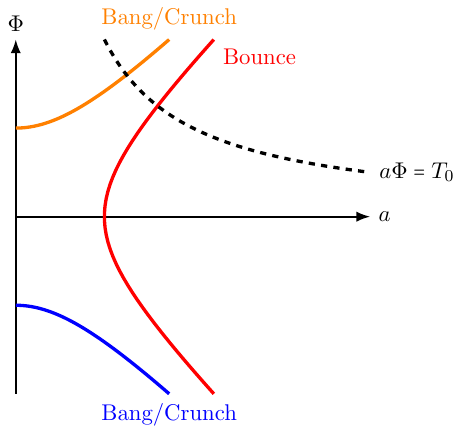}
    \caption{The $(\Phi,a)$ half-plane is the configuration space of the mini-superspace theory. Representative classical trajectories are indicated, and time can increase in either direction. Bounce solutions are in red, big bang or crunch solutions (depending on the direction) with positive or negative dilaton are shown in yellow or blue respectively. We study a Klein-Gordon norm associated to gauge-fixing the Hamiltonian constraint to a dashed surface of constant $a\Phi$. This does not capture the blue trajectories, and will give different signs for red/yellow trajectories depending on the direction of increasing time.}
    \label{fig:aPhiPlane}
\end{figure}

To illustrate these ideas, we will look at the Klein-Gordon norm on slices $\Sigma$ of constant $a\Phi$. This is convenient because the symmetry $G_3$ of the Hamiltonian leaves these slices invariant. It also facilitates comparison with previous discussion of the Klein-Gordon norm in JT \cite{Maldacena:2019cbz,Cotler:2024xzz}.

To do this, we first pass to new coordinates $T,\sigma$ on the $\Phi>0$ region of mini-superspace, chosen to make the symmetry $G_3=ap-\Phi k$ manifest as translations  $-i\partial_\sigma$ in $\sigma$:
\begin{equation}
    T=a\Phi, \qquad \sigma = \frac{1}{2}\log\frac{a}{\Phi}.
\end{equation}
(This differs from \cite{Maldacena:2019cbz,Cotler:2024xzz} by a factor of $2$ in the definition of $\sigma$.) In these coordinates, the metric on mini-superspace becomes $ds ^2=-2d\Phi da = -\frac{dT^2}{2T}+2 T d\sigma^2$, and the Hamiltonian acts as
\begin{equation}
    H =  T\partial_T^2 + \partial_T - \frac{1}{4T}\partial_\sigma^2 +T.
\end{equation}
With this, the Klein-Gordon inner product on a slice of constant $T>0$ is
\begin{equation}
   \li \psi_2|\psi_1\ri_{\mathrm{KG}(T)} =  i \int d\sigma T\left( \psi_2^* \partial_T \psi_1- (\partial_T \psi_2^*) \psi_1 \right).
\end{equation}
We can use the basis of solutions to the Wheeler-DeWitt equation given in section \ref{sssec:G3}, diagonalising the $G_3$ symmetry:
\begin{align}
    \psi_{\lambda,0}(T,\sigma) &=  \Gamma \left(\tfrac{1 +i\lambda}{2}\right) 2^{\frac{i\lambda-1}{2}}  J_{\frac{i\lambda}{2}}(T) \frac{e^{i\lambda\sigma}}{\sqrt{2\pi}} , \\
    \psi_{\lambda,+}(T,\sigma) &=\frac{i }{2}e^{-\frac{\pi}{2}\lambda}\Gamma \left(\tfrac{1+i\lambda}{2}\right)  2^{\frac{i\lambda-1}{2}}  H^{(1)}_{\frac{i\lambda}{2} }(T) \frac{e^{i\lambda\sigma}}{\sqrt{2\pi}}, \\
    \psi_{\lambda,-}(T,\sigma) &=-\frac{i }{2}e^{\frac{\pi}{2}\lambda}\Gamma \left(\tfrac{1+i\lambda}{2}\right)  2^{\frac{i\lambda-1}{2}}  H^{(2)}_{\frac{i\lambda}{2} }(T) \frac{e^{i\lambda\sigma}}{\sqrt{2\pi}}.
\end{align}
Recall that these give an orthonormal basis with respect to the physical group-average inner product, with inner products $\delta(\lambda-\lambda')$ in each ($0,+,-$) sector \eqref{eq:G3GA}.

Before calculating the Klein-Gordon inner product, let's consider our classical expectations based on the preceding discussion. For the bounce solutions ($0$ sector) the dilaton profile is $\Phi = \tilde{\varphi} \sinh t$, and $\lambda = G_3= - b \tilde{\varphi}$. This solution crosses the gauge-slice in the direction of increasing $T$ for $\lambda<0$ (so our Klein-Gordon norm should agree with the physical inner product), and decreasing $T$ for $\lambda>0$ (so we should find a sign discrepancy). The bang solutions (in the $-$ sector) have dilaton profile $\Phi = \varphi \cosh t$ ($t>0$) and $\lambda = \beta \varphi$, so for $\lambda>0$ they cross the constant $T>0$ slice in the increasing direction and should give positive Klein-Gordon norm, but in the $\lambda<0$ case the dilaton is always negative so the solution never satisfies the gauge condition and we expect zero norm. Similarly, the crunch solutions for $\lambda<0$ should give negative norm, and zero Klein-Gordon norm for $\lambda>0$.

Now the calculation of the Klein-Gordon inner products is straightforward, giving results that are diagonal in $G_3$ and independent of $T$ (for $T>0$) as expected. We can summarise the result in a $3\times 3$ matrix with rows/columns labelled by $0,+,-$:
\begin{equation}\label{eq:KGmatrix}
    \li \psi_{\lambda,0/+/-}|\psi_{\lambda,0/+/-}\ri_{\mathrm{KG}(T)} = \begin{pmatrix}
        -\tanh\tfrac{\pi \lambda}{2} & -\tfrac{i}{2}\sech\tfrac{\pi \lambda}{2} & -\tfrac{i}{2}\sech\tfrac{\pi \lambda}{2} \\
         \tfrac{i}{2}\sech\tfrac{\pi \lambda}{2} & -\frac{1}{1+e^{\pi \lambda}}& 0\\
         \tfrac{i}{2}\sech\tfrac{\pi \lambda}{2} & 0 & \frac{1}{1+e^{\pi \lambda}}
    \end{pmatrix}\delta(\lambda-\lambda').
\end{equation}
The first thing to note is that for $\lambda\to \pm \infty$, this matrix becomes diagonal with entries $(-1,0,0)$ for $\lambda\to+\infty$ or $(+1,-1,+1)$ for $\lambda\to-\infty$. Comparing to the above classical discussion (and recalling that we have normalised these states in the group-average inner product), we reproduce expectations from the gauge-fixing intuition.

However, it is important to note that this is no longer true for finite $\lambda$. Nor is there any different basis where such a simple gauge-fixing interpretation applies exactly: the eigenvalues of the above matrix are not integers. But large $\lambda$ can be thought of as a classical limit (with $\lambda\propto \hbar^{-1}$), in which the corrections are non-perturbatively small.

We could also study the Klein-Gordon form on a different hypersurface, for example constant $\Phi$ as in \cite{Nanda:2023wne}. Similar considerations apply, though there are additional subtleties  from the $a=0$ boundary. In particular, the Klein-Gordon inner product is not conserved as we change $\Phi$, since the gauge slice intersects different gauge orbits. This makes the results harder to interpret so we do not explore this further.
 
\subsection{Klein-Gordon S-matrix}

As one more point of comparison, we can look at overlaps of asymptotic states using the Klein-Gordon norm. To do this, we first work out how to write the scattering states $|\phi_{\mathrm{in/out}}\ri$ as wavefunctions in the $|\psi_{\lambda,0/+/-}\ri$ basis. For this we can use the physical group-average inner product to evaluate overlaps (from integrals of $\psi_{0,+,-}$ wavefunctions), since the basis of $\lambda$ eigenstates is orthonormal with respect to this inner product. We obtain
\begin{equation*}
    \begin{gathered}
        \li \psi_{\lambda,0}|\phi_{\mathrm{in}}\ri = \frac{1}{4\pi} \left(\frac{2i}{\phi_{\mathrm{in}}}\right)^\frac{1-i\lambda}{2} \Gamma\left(\tfrac{1-i\lambda}{2}\right), \quad \li \psi_{\lambda,+}|\phi_{\mathrm{in}}\ri = \frac{1}{4\pi} \left(-\frac{2i}{\phi_{\mathrm{in}}}\right)^\frac{1-i\lambda}{2} \Gamma\left(\tfrac{1-i\lambda}{2}\right), \quad \li \psi_{\lambda,-}|\phi_{\mathrm{in}}\ri = 0, \\
        \li \psi_{\lambda,0}|\phi_{\mathrm{out}}\ri = \frac{1}{4\pi} \left(-\frac{2i}{\phi_{\mathrm{out}}}\right)^\frac{1-i\lambda}{2} \Gamma\left(\tfrac{1-i\lambda}{2}\right), \quad \li \psi_{\lambda,+}|\phi_{\mathrm{out}}\ri=0 , \quad \li \psi_{\lambda,-}|\phi_{\mathrm{out}}\ri=\frac{1}{4\pi} \left(\frac{2i}{\phi_{\mathrm{out}}}\right)^\frac{1-i\lambda}{2} \Gamma\left(\tfrac{1-i\lambda}{2}\right).
    \end{gathered}
\end{equation*}
Some care should be taken with phases here (taking definitions of powers with branch cuts along the negative real axis).

These inner products give us the  components of in- and out-states in the $G_3$ basis, so to find the Klein-Gordon overlaps we combine these vectors of components with  the matrix \eqref{eq:KGmatrix} and integrate over $\lambda$. For example, when we calculate $\li \phi_\mathrm{out}|\phi_\mathrm{in}\ri_{\mathrm{KG}(T)}$ for $\phi_\mathrm{in}<0$  and $\phi_\mathrm{out}>0$ we arrive at the integral
\begin{equation}
    \frac{1}{4\pi\sqrt{-\phi_\mathrm{in}\phi_\mathrm{out}}}\int \left(\frac{-\phi_\mathrm{in}}{\phi_\mathrm{out}}\right)^{\frac{i\lambda}{2}}\frac{d\lambda}{1+e^{\pi \lambda}},
\end{equation}
which we can evaluate using $\int \frac{e^{-i\lambda x}}{1+e^{\pi\lambda}}d\lambda = \frac{i}{\sinh x+i\epsilon}$. The result of this calculation for the S-matrix $\li \phi_\mathrm{out}|\phi_\mathrm{in}\ri_{\mathrm{KG}(T)}$ is
\begin{equation}
    \li \phi_\mathrm{out}|\phi_\mathrm{in}\ri_{\mathrm{KG}(T)} = \begin{cases}
        \frac{i}{2\pi(\phi_\mathrm{in}+\phi_\mathrm{out}+i\epsilon)} &  \phi_\mathrm{in}<0,\quad \phi_\mathrm{out}>0 \\
        -\frac{i}{2\pi(\phi_\mathrm{in}+\phi_\mathrm{out}+i\epsilon)} & \phi_\mathrm{in}>0,\quad \phi_\mathrm{out}<0 \\
        0 & \phi_\mathrm{in}>0,\; \phi_\mathrm{out}>0 \text{ or } \phi_\mathrm{in}<0,\; \phi_\mathrm{out}<0.
    \end{cases}
\end{equation}
This agrees precisely with what we would have guessed from the general gauge-fixing discussion. In the first case we are considering configurations with large negative $\Phi a$ in the past and large positive $\Phi a$ in the future, so constant $T =\Phi a$ constitutes a good gauge, and the result agrees precisely with the correct result \eqref{eq:S-matrix} from the physical inner product. In the second case all possible configurations cross in the opposite direction, so the result disagrees with \eqref{eq:S-matrix} by a sign. When $\phi_\mathrm{in}$ and $\phi_\mathrm{out}$ have the same sign, all configurations begin and end on the same side of the constant $T$ gauge slice, so the corresponding Klein-Gordon inner product gives zero.

It is interesting to note that even when this Klein-Gordon S-matrix  agrees with the group-average inner product (e.g., for $\phi_\mathrm{in}<0$ and $\phi_\mathrm{out}>0$), it does not receive contributions only from the bounce sector (the top left $00$ component of the matrix \eqref{eq:KGmatrix}). There are also off-diagonal contributions mixing bounce and bang/crunch sectors (though they are small in a semi-classical regime).

We can also calculate Klein-Gordon overlaps of a pair of in- or out-states, but the results are not so simple or easy to interpret. This should not be surprising, since the Klein-Gordon inner product of in-states will have contributions from the distant future for some histories but not from others.

\subsection{Geodesic gauge and other generalisations of Klein-Gordon inner products}\label{ssec:kIP}

While the Klein-Gordon inner product is best known, our above discussion in terms of gauge-fixing makes it clear that one can greatly generalise to give different inner products on invariants that may be more useful. To do this, we can take a gauge-fixing function $F$ which is a more general Hermitian operator, not just a function of mini-superspace configuration space variables (i.e., it can depend on momenta).

One particularly interesting possibility (particularly in higher-dimensional theories of gravity) fixes to a slice of vanishing mean curvature, $\Tr \mathcal{K}=0$: see \cite{witten2023notecanonicalformalismgravity} and references therein for a detailed discussion. A simple analogy in our JT model is to choose $F=-k$, fixing to a geodesic slice (the sign choice simply being a useful convention). This gives us a gauge-fixing operator $\kappa_k=i[H,\Theta(F)]=a\delta(k)$, fixing $k$ to zero along with the (positive) measure factor $a$, being particularly simple because $a=i[H,F]$ commutes with $F$.  The resulting overlap $\li \cdot|\cdot\ri_k$ agrees precisely with the group-average result restricted to the bounce sector \eqref{eq:GAbounce}, in which case there exists a $k=0$ slice satisfying the gauge condition (and $k$ is always decreasing):
\begin{equation}
    \li \psi'|\psi\ri_k = \int_0^\infty b db \psi'_0(b)^*\psi_0(b).
\end{equation}
The measure factor in this gauge-fixing is one way to understand the familiar $bdb$ measure when we glue along geodesics in the path integral (for example, in \cite{saad2019jtgravitymatrixintegral}). However, this is not a good gauge for the bounce and crunch sectors, since the corresponding spacetimes contain no closed geodesics.

More generally, a particularly simple family of gauge choices is given by taking any function $F(a,k)$. For example we can take $F=k_0-k$ for some constant $k_0$, or $F=\mathcal{K}-\frac{k}{a}$ to fix to slices of constant extrinsic curvature $\mathcal{K}$ (this is  the `York time', when regarded as a function parameterising a foliation of Cauchy surfaces). These are simple because the constraint $H$ is linear in the momenta $p,\Phi$ conjugate to the chosen variables $a,k$ (as is typical for ordinary gauge theories). We can analyse this very simply by passing to the Rindler coordinates, writing $F(u,b)$ (and similarly using Milne coordinates in bang/crunch regions). Then $H=i\partial_u$, so we get a `Fadeev-Popov' factor $i[H,F]=-\frac{\partial F}{\partial u}$ which commutes with the gauge-fixing operator $\delta(F)$. For fixed $b$ we can rewrite $\delta(F)$ as a sum of terms proportional to $\delta(u-u_0(b))$ where $u=u_0$ give the values at which $F=0$. These each come with factors $|\partial_uF|^{-1}$ (assuming this derivative is non-zero) which cancel the measure up to a sign:
\begin{equation}
    \kappa_F = \sum_{F=0} \sgn(-\partial_u F) \delta(u-u_0(b)).
\end{equation}
This precisely realises the gauge-fixing discussion above. $F$ defines a good gauge for a trajectory which crosses the $F=0$ slice once in the increasing direction (or more generally begins at negative $F$ and ends at positive $F$). For states supported only on such orbits, the $\kappa_F$ inner product will agree exactly with the group-average inner product: $\eta \kappa_F$ acts as the identity on these states, so $\kappa_F$ is playing the role of an inverse to $\eta$.

As an example, choosing  $F=\mathcal{K}-\frac{k}{a}$ gives us $\kappa_\mathcal{K} = \left(1-\frac{k^2}{a^2}\right)\delta\left(\mathcal{K}-\frac{k}{a}\right) = a(1-\mathcal{K}^2)\delta(k-\mathcal{K}a)$. This gives us an inner product
\begin{equation}
    \li \psi|\psi\ri_\mathcal{K} = \int_0^\infty da \, a (1-\mathcal{K}^2) |\psi(a,k=\mathcal{K}a)|^2.
\end{equation}
For $-1<\mathcal{K}<1$ this is positive, picking up the bounce sector and agreeing with the group-average inner product there. For $\mathcal{K}>1$ or $\mathcal{K}<-1$ we instead pick up the crunch or bang component, and get \emph{minus} the group-average inner product, since trajectories cross the $F=0$ slice in the direction of decreasing $F$. By adapting the choice of $\mathcal{K}$ to the state of interest (or combining multiple possibilities) and choosing appropriate signs, we can get an inner product on invariants which agrees exactly with the group-averaging procedure (though no single simple choice captures all states at once).

\section{Lifting to the full two-dimensional theory}\label{sec:2D}

We now revisit the connection of the mini-superspace model we've been studying to the full theory of dS JT gravity, using some of the ideas we have encountered along the way. Like the mini-superspace theory, this is most straightforward in the $(a,k)$ basis, and we can give a relatively complete and explicit account. But if we try to translate to the $(a,\Phi)$ variables --- in principle, just a Fourier transform from $k$ to $\Phi$ --- we will see reasons to expect the theory to be very complicated. However, there is one regime where it is possible, namely for large universes with large dilaton where we connect the Fourier transform with the `Schwarzian' description. We discuss each of these in turn.

\subsection{Exact reduction to  mini-superspace in $k'=a'=0$ gauge}

Now, wavefunctions are functionals $\Psi[k(\theta),a(\theta)]$ depending on the spatial metric $a(\theta)$ and $k(\theta)$, which is $a$ times the extrinsic curvature. 
We have already commented on the Wheeler-DeWitt invariant states in section \ref{ssec:akStates}. For fixed $a(\theta),k(\theta)$ we expect there to be a single geometry of constant curvature $\mathcal{R}=2$ which embeds a curve with the corresponding length and  extrinsic curvature, and the invariant state wavefunctions depend only on this embedding geometry. We can recover our mini-superspace truncation by restricting to these wavefunctions evaluated at $a'(\theta)=k'(\theta)=0$. This excludes the $n\times\mathrm{dS}_2$ geometries but does not otherwise lose anything since the wavefunctions at other values  can (in principle) be constructed exactly from the mini-superspace slice.

Alternatively, we can think of the mini-superspace states as co-invariants represented by wavefunctions proportional to a functional delta-functions $\delta(a'(\theta))\delta(k'(\theta))$ setting $a,k$ to be independent of $\theta$. These two possibilities (invariant or co-invariant with respect to spatially-dependent modes of the Hamiltonian and momentum constraints $\mathcal{H},\mathcal{P}$) are not identical: we get from co-invariants to invariants  by the group-averaging rigging map, or inversely from invariants to co-invariants by one of the gauge-fixing maps $\kappa$  discussed in the previous section. This translation involves a measure factor (like the factor of $b$ in \eqref{eq:GAbounce}), which will differ from the mini-superspace factor from including the spatially varying constraints. This will also give us the inner product, allowing us to more precisely identify the relation between the mini-superspace and full wavefunction.

For this, consider the partial gauge-fixing $a'=k'=0$ which takes us from the full model the the mini-superspace. The associated Fadeev-Popov measure is the determinant of a matrix of Poisson brackets (or commutators) with the constraints, $\{(\mathcal{P},\mathcal{H}),(a',k')\}$ (where we include only the spatially varying parts, or the non-zero Fourier modes).\footnote{To derive these result, we can use $\{\mathcal{P}[\xi],a\} = -(a \xi)'$,  $\{\mathcal{P}[\xi],k\} = -(k \xi)'$, $\{\mathcal{H}[\eta],a\} = k \eta$, $\{\mathcal{H}[\eta],k\} = a \eta+ a^{-1} \eta''$, where the notation is the same as in \eqref{eq:algebra}.} This is diagonal in the Fourier basis (modes $\propto e^{im\theta}$ for non-zero integer $m$),
\begin{equation}\label{eq:muFP}
    \mu_\mathrm{FP} = \prod_{m\neq 0} \det \begin{pmatrix}
    m^2 a & m^2 k \\ im k & im a^{-1}(a^2-m^2) 
\end{pmatrix}= \mathcal{N} \frac{\sin^2\left(\pi\sqrt{a^2-k^2}\right)}{\pi^2(a^2-k^2)},
\end{equation}
where the rows of the matrix corresponds to $\mathcal{P}$ and  $\mathcal{H}$, and the columns to $a'$ and $k'$. $\mathcal{N}$ is a divergent (but field-independent) normalisation factor; we should regulate this in some consistent way, but the result only affects the overall normalisation in any case. Combining this with the mini-superspace calculations above (to account for the $m=0$ modes) in the $k=0$ gauge-fixing (as a concrete example), we get a map from an invariant wavefunction to a gauge-fixed co-invariant,
\begin{equation}\label{eq:kappakFull}
    \Psi[a,k] \xrightarrow{\quad\kappa_k\quad } \delta(k)\delta(a') \frac{\sin^2 (\pi b)}{\pi^2 b}\Psi[a(\theta)=b,k(\theta)=0].
\end{equation}
This also immediately gives us a corresponding inner product of invariant states,
\begin{equation}
    \li \Psi|\Psi \ri_k =  \int_0^\infty  |\Psi[b,0]|^2  \frac{\sin^2 (\pi b)}{\pi^2 b}db,
\end{equation}
where we should recall that this is only the contribution to the inner product from the bounce sector where the $k=0$ gauge is good. The full answer will include similar integrals with measure $\frac{\sinh^2 (\pi \beta)}{\pi^2 \beta}d\beta$ from bang and crunch sectors (as in \eqref{eq:GAinv}), as well as from $n\times\mathrm{dS}_2$ sectors.

While the  change in the measure makes it appear that the spatial dependence modifies the mini-superspace theory, we expect physical wavefunctions (e.g., the asymptotic scattering wavefunctions in section \ref{ssec:scatStates}) to pick up compensating factors of $\frac{\pi b}{\sin(\pi b)}$ from fluctuations of spatially varying modes. This means that we recover the mini-superspace by rescaling wavefunctions and simultaneously rescaling the measure in the physical inner product to leave all physical amplitudes invariant. Evidence for this comes from studying the AdS (Euclidean) JT theory \cite{Goel:2020yxl}, where precisely such factors appeared from a cylinder topology path integral  with constant extrinsic curvature boundary conditions (see also discussions of torsion in \cite{Stanford:2019vob}). We do not understand this completely, so a careful analysis would be interesting.

A full treatment of the 2D theory should also account carefully for the  $n\times\mathrm{dS}_2$ sectors, which the $k'=0$ gauge does not capture. This should be manageable by using the gauge condition in \eqref{eq:ksqGF}, since this involves only the geometric fields $a,k$. A potential subtlety is that the $n\times\mathrm{dS}_2$ solutions are invariant under a non-compact one parameter symmetry, as opposed to the compact $U(1)$ of rotations in the other sectors. This would lead to a divergence in the group average integral, so we would need to exclude this mode to get a sensible inner product.

\subsection{Passing to $(a,\Phi)$ variables}

For our mini-superspace description, passing to the $(a,\Phi)$ representation was a straightforward Fourier transform with respect to $k$. But in the full theory, this is only transforming the zero-mode of $k$ (fixing $k'=0$), so the mini-superspace $\Phi(t)$ corresponds to the zero-mode (or spatial average) $\int \frac{d\theta}{2\pi}\Phi(t,\theta)$ of the dilaton in the full theory. To get the true wavefunctions $\Psi[\Phi(\theta),a]$  we need to transform all modes (even if we only want to evaluate at constant $\Phi$), doing a functional Fourier transform
\begin{equation}\label{eq:FT}
    \Psi[\Phi,a]= \int \mathcal{D}k(\theta) e^{i \int \frac{d\theta}{2\pi}k(\theta)\Phi(\theta)} \Psi[k,a].
\end{equation}
This requires integrating over all possible $k(\theta)$.

For co-invariant states with a delta function setting $k'=0$ (such as the right side of \eqref{eq:kappakFull}), the non-zero mode integrals are trivial, but they tell us that the transformed wavefunction depends on $\Phi$ only through its zero mode: spatial variations of the dilaton will have completely uncontrolled quantum variance. This is the way to interpret the states we have been discussing in the mini-superspace.

For invariant states, we expect this transform to $\Psi[\Phi(\theta),a]$ to be very complicated. The basic reason is that (even if we only want to evaluate for constant $\Phi(\theta)$) we need to integrate over all possible $k(\theta)$. Determining the wavefunction $\Psi[k(\theta),a]$ in this integral requires us to first identify which geometry contains a closed curve with specified proper length and extrinsic curvature (as a function of $\theta$), which is a complicated non-local functional.\footnote{In some ways this is not as bad as in Euclidean signature where we may want to also exclude functions $k(\theta)$ which lead to self-intersecting curves \cite{Stanford:2020qhm}, which cannot happen for spacelike closed curves in a 2D Lorentzian geometry.} It would be interesting to try to solve this problem, though we are somewhat pessimistic that it will be tractable in general. Nonetheless, it is possible in certain limits, one of which we describe below.

Despite this apparent challenge, previous work \cite{Nanda:2023wne} (following AdS calculations \cite{Iliesiu:2020zld} and much earlier work \cite{Henneaux:1985nw,Louis-Martinez:1993bge}) claimed a simple general solution to the Wheeler-DeWitt equation for wavefunctions $\Psi[\Phi(\theta),a]$ depending on the dilaton and size of the universe. But there are several reasons to question their approach, which we now briefly explain.

The first idea is to use the constancy of the quantity $M$ given in \eqref{eq:Phiconst}, written in \eqref{eq:Phiconst2} in our canonical variables. This is analogous to looking for solutions with definite $G_2$ in our mini-superspace model (section \ref{sssec:G2}). But to do this, they do not directly quantise $M$ as written. Instead, they first solve classically for momentum $p$ in terms of $a$, $\Phi$ and the constant $M$, and then similarly solve constraints for $k$ (eliminating $p$) to write
\begin{equation}
    p = \pm \sqrt{\Phi^2 + a^{-2}(\Phi')^2-M}, \qquad k = \pm\frac{a\Phi -a^{-2}a'\Phi'+a^{-1}\Phi''}{\sqrt{\Phi^2 + a^{-2}(\Phi')^2-M}}.
\end{equation}
These two equations are classically equivalent to the constraints and the specification of the constant $M$. The second ideas is to quantise not the original constraints, but instead these relations, imposing them as first-order differential equations (using $p=-i\frac{\delta}{\delta a}$, $k=-i\frac{\delta}{\delta \Phi}$) to define Wheeler-DeWitt wavefunctions.

This approach is of course \emph{not} equivalent to solving the original constraints and diagonalising $M$ as a second-order differential operator (quadratic in $p$), for the same reason that eigenfunctions of $\frac{p^2}{2m}+V(q)$ are not obtained by solving $p|\psi\rangle=\pm\sqrt{2m(E-V(q))}|\psi\rangle$ (unless $V$ is constant so commutes with $p$). One might argue that this is simply a different choice for definition of the quantum theory, perhaps arising from a different operator ordering. But this definition is not particularly natural or well-motivated (except for being a choice that makes the Wheeler-DeWitt equations in the $a,\Phi$ basis easy to solve). One issue is that the corresponding operator ordering for the constraints contains explicit dependence on the constant $M$, which is not very satisfactory. But more importantly, this choice does not agree with the usual path-integral quantisation of JT, which makes explicit use of linearity in the dilaton to treat $\Phi$ as a Lagrange multiplier imposing constant curvature. As we have already mentioned, the analogous statement in the canonical formalism is that the constraints are linear in $\Phi$ and $p$, and that they become first-order (functional) differential operators acting on wavefunctions  $\Psi[k,a]$. For this reason we prefer the obvious quantum definition of the constraints (which in our variables do not na\"ively suffer from operator ordering ambiguities in our variables, aside from a possible constant in $\mathcal{P}$) as the natural choice for quantising the theory, and the choice which agrees with the path integral definition restricted to constant curvature manifolds.\footnote{Another potential source of guidance could be the BRST formalism, where the requirement for a Hermitian nilpotent BRST charge is rather constraining (particularly in the gravitational case with field-dependent structure constants).}

\subsection{Connection to the Schwarzian}

Though we have been rather pessimistic about a useful description of wavefunctions $\Psi[\Phi,a]$, from the existing work on the path integral \cite{Maldacena:2019cbz,Cotler_2020} we know that there is at least one regime where this is tractable: namely for describing wavefunctions in the asymptotic regime of large $a$ and $\Phi$, corresponding to large universes near $\scri_\pm$. Here we sketch how this can be seen from our canonical approach.

For definiteness, we will concentrate on the bounce sector part of the wavefunction, taking a wavefunction $\Psi[k,a]$ which solves the Wheeler-DeWitt equation and evaluates to $\psi_0(b)$  on bounce geometries (so $\Psi[k=0,a(\theta)=b] = \psi_0(b)$). Other sectors are similar. We wish to evaluate its Fourier transform $\Psi[\Phi(\theta),a(\theta)]$ from \eqref{eq:FT} near the asymptotic future region $\scri_+$, which means taking a limit $a\to\infty$  and $\Phi\to\infty$ with fixed ratio $\frac{\Phi}{a}$ (we take $a$ constant but allow $\Phi(\theta)$ to be a function of $\theta$; non-constant $a$ follows immediately from spatial diffeomorphisms).

The first thing we must do is identify which $k(\theta)$ contribute significantly to the Fourier integral \eqref{eq:FT}. At large $\Phi$ the phase $e^{i\int k\Phi}$ is highly oscillatory so we should be looking near points of stationary phase, which are essentially classical solutions for $k$. Since $k=-\dot{a}$ and $a$ grows exponentially at late times $a\propto e^t$, our stationary points will have $k\sim -a$. So the Fourier transform will be dominated by $k(\theta)\approx -a$. For the bounce solutions we would also pick up a contribution from $k\approx + a$ corresponding to the distant past: we will focus on the `positive frequency part' of the wavefunction (which we could pick out using the scattering state definition of \ref{ssec:scatStates} for example).

So, we are looking for Cauchy surfaces at late times in the bounce sector metric as $ds^2 = -dt^2+b^2\cosh^2 t\, d\vartheta^2$, where we use spatial coordinate $\vartheta$ to distinguish from the usual angular coordinate $\theta$ on a desired embedded spatial circle. If the surface does not vary too much in the time direction ($|t'(\theta)|\ll b \cosh t\, \vartheta'(\theta)$, as required so that $k(\theta)\approx -a$), then the condition that $a d\theta$ is the induced proper length gives us the $t$ coordinate of the surface in terms of the diffeomorphism $\vartheta(\theta)$:
\begin{equation}
    \frac{d \vartheta}{d\theta} \sim \frac{2a}{b}e^{-t}.
\end{equation}
We can then compute the extrinsic curvature of such a surface at large $a$ (and fixed $\vartheta(\theta)$), finding
\begin{equation}
    \mathcal{K} = 1  -\frac{1}{a^2}\left[\{\vartheta,\theta\}+\frac{b^2}{2}\right] + O(a^{-4}) \implies k(\theta)= - a+ a^{-1}\left[\{\vartheta,\theta\}+\frac{b^2}{2}\left(\frac{d\vartheta}{d\theta}\right)^2\right] + \cdots,
\end{equation}
where $\{\vartheta,\theta\}$ is the Schwarzian derivative.  This gives us a map taking a number $b^2$ and a diffeomorphism $\vartheta(\theta)$ to a function $k(\theta)$ (or more precisely the function $a(k(\theta)+a)$ in the $a\to\infty$ limit). This map is in fact invertible up to the $U(1)$ of adding a constant to $\vartheta$,\footnote{One way to understand this uses the machinery of coadjoint orbits of the Virasoro group. The combination in the square brackets is the coadjoint action of the Virasoro group (labelled by the diffeomorphism $\vartheta$ and a central charge) on a constant function proportional to $b^2$. So this constant labels the coadjoint orbit which the function $a(k(\theta)+a)$  belongs to, and $\vartheta$ the diffeomorphism required to map to the constant function representative of the orbit \cite{Stanford:2017thb}.} so we can exchange the integral over functions  $k(\theta)$ for an ordinary integral over $b$ along with an integral over $\vartheta\in \operatorname{Diff}(S^1)/U(1)$. After doing this (without taking particular care over the measure here), the functional Fourier transform \eqref{eq:FT} becomes
\begin{equation}\label{eq:Schw}
    \Psi[\Phi(\theta),a] \sim  e^{-i a \int \frac{d\theta}{2\pi}\Phi(\theta)}\int_0^\infty db \int \frac{\mathcal{D}\vartheta(\theta)}{U(1)} \exp\left[  i \int \frac{d\theta}{2\pi}  \left(\{\vartheta,\theta\}+\tfrac{b^2}{2}
    \left(\tfrac{d\vartheta}{d\theta}\right)^2\right)\frac{\Phi(\theta)}{a} \right] \psi_0(b).
\end{equation}
In the limit of interest ($\frac{\Phi(\theta)}{a}$ fixed when we take $a\to\infty$), this Fourier transform becomes precisely the Schwarzian path integral: specifically, it reproduces the result of \cite{Cotler:2023eza} which allows for non-constant dilaton.

This holds for generic values of $b$, and there is a similar result taking $b^2\rightarrow-\beta^2$ for the big bang solutions (and big crunch if we focus on the past asymptotic region). Our wavefunctions are thus computed by path integrals \eqref{eq:Schw} containing  a Schwarzian theory with imaginary weighting in the action (or in terms of the Schwarzian coupling $g$, we have imaginary $g^2$). Apart from this important imaginary weighting, these are the same theories which appear in Euclidean AdS JT for the trumpet \cite{saad2019jtgravitymatrixintegral} (corresponding to the bang/crunch sectors) or the disk with conical defects  \cite{Mertens:2019tcm} (or excesses for $b>1$) for the bounce sectors. Integer values of $b$ are special (with extra zero-modes appearing in the Schwarzian action), signalling the appearance of additional symmetries. An example of this appears for the Hartle-Hawking state corresponding to $b=1$ --- see comments in section \ref{ssec:HH} --- which makes contact with the Schwarzian theory on the hyperbolic disc for the AdS theory.

\subsection{A new Schwarzian theory for $n\times \mathrm{dS}_2$ geometries}

It remains to examine the $n\times \mathrm{dS}_2$ sector contributions. We can import much of the discussion above directly to this case, since we can write the metric of these spacetimes just like a bounce metric $ds^2 = -dt^2 + n^2 \cosh^2 t d\vartheta^2$, except with non-trivial identifications. We described this earlier \eqref{eq:nAdSsol} using a single patch with $0<\vartheta<2\pi$ and gluing its edges $(t,\vartheta=0)\sim(t+\beta,\vartheta=2\pi)$, but we will here find an alternative  useful, which allows $\vartheta$ to take real values but with identifications. Specifically, a point $(t,\vartheta)$ in the fundamental domain $0<\vartheta<2\pi$ is identified with a point in the region $2\pi k <\vartheta<2\pi(k+1)$, specified by following the flow of the diffeomorphism $\cos(n\vartheta)\partial_t - \frac{1}{n}\sin(n\vartheta)\tanh t \,\partial_\vartheta$ a distance $k\beta$ starting at $(t,\vartheta+2\pi k)$. We are particularly interested in the action near $\scri_+$ (large $t$), where this finite diffeomorphism acts by mapping $\tan\left(\frac{n \vartheta}{2}\right)\mapsto e^{-\beta}\tan\left(\frac{n \vartheta}{2}\right)$. This identification is inherited by the map $\vartheta(\theta)$ that defines the embedded Cauchy surface as above.

The upshot is that we land on a Schwarzian theory with the same action as in \eqref{eq:Schw} (with $b=n$), but with different quasi-periodic boundary conditions on $\vartheta$:
\begin{gather}
    I = \int \frac{d\theta}{2\pi}  \left(\{\vartheta,\theta\}+\tfrac{n^2}{2}
    \left(\tfrac{d\vartheta}{d\theta}\right)^2\right)\Phi_r(\theta), \qquad
    \tan\left(\tfrac{n }{2}\vartheta(\theta+2\pi)\right) = e^{-\beta} \tan\left(\tfrac{n }{2}\vartheta(\theta)\right),
\end{gather}
where $\Phi_r=\frac{\Phi}{a}$ is the `renormalised dilaton'. We can alternatively write this in terms of $f(\theta) = \tan\left(\tfrac{n }{2}\vartheta(\theta)\right)$, in terms of which the action and boundary conditions simply become
\begin{equation}
    I = \int \frac{d\theta}{2\pi} \{f,\theta\}\Phi_r(\theta), \qquad
    f(\theta+2\pi) = e^{-\beta} f(\theta).
\end{equation}
However, we also need to specify a topological condition on $f$ capturing the requirement that $\vartheta$ wraps the circle once when $\theta$ increases by $2\pi$. One way to state this is that $f$ must be an increasing function with $n$ simple poles in each $2\pi$ period (corresponding to the points where $\frac{n}{\pi}\vartheta$ is  an odd integer). This means, for example, that $f(\theta) = e^{-\frac{\beta}{2\pi}\theta}$ is not an admissible configuration. The isometry of the spacetime leaves its mark on this Schwarzian theory as invariance under an overall rescaling of $f$ (which gets enhanced to $PSL(2,\RR)$ only for $\beta=0$), which is a redundancy. Our path integral should therefore run over a $\operatorname{Diff}(S^1)/\RR$ coset: if we choose some fiducual function $f=f_0$ obeying the boundary conditions then any other $f$ is obtained by composing $f_0$ with a diffeomorphism (giving the $\operatorname{Diff}(S^1)$), while the quotient by $\RR$ is analogous to the $U(1)$ rotation above but is now non-compact.

This is an entirely novel Schwarzian theory (as far as we know)  with some rather curious features. For example, it has no real classical solutions  for  constant $\Phi_r$ (associated with the fact that the classical dilaton solutions near $\scri_+$ change sign, going as $ e^t\cos(n\vartheta)$). We also expect that it only makes sense as a `Lorentzian' theory (integrating $e^{iI}$ rather than the `Euclidean' $e^{-I}$), since the action is not bounded from below. An alternative characterisation uses the association between the Schwarzian path integral and the coadjoint orbits of the Virasoro group \cite{Stanford:2017thb}: our new theory is associated to the orbits labelled $\operatorname{Diff}(S^1)/T_{(n,\Delta)}$ in the classification of \cite{Witten:1987ty}. It appears that different solutions to de Sitter JT gravity give rise to a Schwarzian theory for every class of coadjoint orbit in that classification!\footnote{The `parabolic' orbits $\tilde{T}_{(n,\pm)}$ should correspond to spacetimes obtained by identifications under diffeomorphisms generated by $\cos(n\vartheta)\partial_t - \frac{1}{n}(\sin(n\vartheta)\tanh t\pm 1) \,\partial_\vartheta$, a `null rotation' in $\mathfrak{so}(1,2)$.} We hope to revisit this theory in future work.\footnote{We would like to thank Douglas Stanford for conversations about this theory.}

\section{Quantum dS JT coupled to matter}\label{sec:Qmatter}

\subsection{The Hilbert space with matter}

A nice feature of JT is that it is not much more complicated to solve the theory with general matter (coupled only to the metric), as described for the classical in section \ref{ssec:clMatter}. In the Hamiltonian formalism, this simplicity appears in the constraints which retain the property of linearity in $\Phi$ and $p$ (so in the quantum theory, they remain first-order differential operators). For example, in the bang/crunch/bounce sectors we can still use the $a'=k'=0$ gauge to reduce to a single Hamiltonian constraint and momentum constraint, which will now take the form
\begin{equation}\label{eq:Hmatter}
    H=-pk+a\Phi+H_\mathrm{matter}(a,k),\qquad P=P_\mathrm{matter}(a,k)
\end{equation}
where $H_\mathrm{matter}(a,k)$ and $P_\mathrm{matter}(a,k)$ are operators in the matter quantum field theory. More precisely,  the Hilbert space of the QFT typically depends on the spacetime geometry, so these operators $H_\mathrm{matter}(a,k)$ and $P_\mathrm{matter}(a,k)$ act on a space $\hilb_\mathrm{matter}(a,k)$ which depends on $a,k$.

Quantisation now proceeds precisely as above, with only minor technical differences. The upshot is that the physical Hilbert space (restricting for now to the sectors captured by the mini-superspace gauge) is a direct integral of matter Hilbert spaces for all possible geometries (with zero momentum):
\begin{equation}
    \hilb_\mathrm{phys} = \int_\oplus \hilb_\mathrm{matter}^{(0)}(b) b db \quad \oplus\quad  \int_\oplus \hilb_\mathrm{matter}^{(+)}(\beta) \beta d\beta \quad \oplus\quad  \int_\oplus \hilb_\mathrm{matter}^{(-)}(\beta) \beta d\beta\,.
\end{equation}
This means that a state is a triple of wavefunctions $\psi_0(b)$, $\psi_\pm(\beta)$ as before. The difference is that
$\psi_0(b)$ is now valued in $\hilb_\mathrm{matter}^{(0)}(b)$ (the Hilbert space of the matter QFT on a bounce geometry of minimal radius $b$ with $P_\mathrm{matter}=0$), and similarly $\hilb_\mathrm{matter}^{(\pm)}(\beta)$ for bang and crunch. The inner product is like \eqref{eq:GAinv}, except that $|\psi_0(b)|^2$ is replaced with the inner product in the matter QFT.

The generalisation to the full theory should be fairly straightforward, simply including the additional $n\times\mathrm{dS}_2$ sectors. The only slight difference is that the matter states in those sectors should be invariant under the static-patch time translation symmetry of the spacetime, which is non-compact. This means we cannot simply project onto the invariant states, but instead we can use the group-average inner product.\footnote{An exception to this comes from dS-invariant matter states, which give non-compact zero modes like the pure JT theory: see comments in section \ref{sec:2D}. This is similar to the usual case of de Sitter (with no dilaton to break the symmetry) \cite{Higuchi_1991,Marolf_2009}.}

\subsection{Example: conformal matter}

To illustrate the general ideas, we describe the example of coupling to a CFT. This allows us to explicitly compute $H_\mathrm{matter}(a,k)$ by Weyl transform to a Lorentzian cylinder. Explicitly, we can write a circularly-symmetric metric as
\begin{equation}
    ds^2 = -N(t)^2dt^2+a(t)^2d\theta^2 = a(t)^2(-d\eta^2+d\theta^2), \qquad  \eta= \int \frac{N(t)}{a(t)} dt,
\end{equation}
where $\eta $ is the conformal time. There is a state-independent contribution to the action on the curved geometry from the conformal anomaly,
\begin{equation}
    I_\mathrm{anom}=-\frac{c}{24\pi}\int d\eta d\theta (\partial_\eta \log a)^2 = -\frac{c}{12} \int \frac{\dot{a}^2}{aN}dt.
\end{equation}
Along with this we have the contribution from the original cylinder Hamiltonian, which gives us $-\int  (\Delta-\frac{c}{12})d\eta =-(\Delta-\frac{c}{12})\int  \frac{N}{a}dt $, where $\Delta = L_0+\bar{L}_0$ is the usual CFT conformal dimension. Combining these, we have a  mini-superspace action (which is still exact),
\begin{equation}
    I = -\int dt  \left[N a \Phi + N^{-1}\dot{\Phi}\dot{a} +\frac{c}{12} \frac{\dot{a}^2}{aN}+ \left(\Delta-\frac{c}{12}\right)\frac{N}{a}\right].
\end{equation}
We can also get the same from analysing the matter stress tensor, demanding that the momentum $\int T_{t\theta}d\theta$ vanish, that the trace is fixed by the conformal anomaly $\Tr T=-\frac{c}{12}$, and conservation $\nabla^\mu T_{\mu \nu}$. The most general solution gives
\begin{equation}
    T_{tt}= \frac{N^2}{a^2}\left(\Delta-\frac{c}{12}\right)-\frac{c}{12}\frac{\dot{a}^2}{a^2}
\end{equation}
for some constant $\Delta$. (We can also have non-zero modes $\frac{N^2}{a^2} L_m e^{im\theta}$ for $m\neq 0$ that will affect the angular-dependent part of the dilaton, but these do not otherwise make a difference in the $k'=0$ gauge.) This is precisely what we get by varying the matter piece in the above action (with $T_{tt}=-\frac{N^2}{a}\frac{\delta I}{\delta N}$).

We can now pass from this Lagrangian description to a Hamiltonian in the usual way, and the Hamiltonian constraint becomes
\begin{equation}
    H = a\Phi - k p +\frac{c}{12}\frac{k^2}{a}+\frac{\Delta-\frac{c}{12}}{a}.
\end{equation}
 Note that the appearance of $\dot{a}$ in the stress tensor modifies the definition of $p$, which  becomes the normal derivative of $\Phi$ (as before) plus a new term $\frac{c}{6}$ times the extrinsic curvature. The constraint remains simple in the $(a,k)$ representation as claimed in \eqref{eq:Hmatter}, since the extra terms from the matter are independent of $\Phi,p$: this follows from matter coupling only to the geometry, not directly to the dilaton.

Before quantising, we briefly note a qualitatively new feature of the dilaton solutions obtained from this constraint (or equivalently \ref{eq:EOMmatter} with the matter stress tensor as a source). For the bang and crunch geometries, $\Phi$ blows up at the $a\to 0$ singularities, going as $\left((\Delta-\frac{c}{12})\beta^{-2}-\frac{c}{12}\right)\log|t|$. Note that this can take either sign if the quantum matter effects (e.g., Casimir energy) are sufficiently strong.



Now, to quantise the theory we first look for solutions to the Wheeler-DeWitt equation. In the bounce sector (for example), this is simplified by working in Rindler coordinates $(b,u)$ where we get
\begin{equation}
  H\psi = \big[i\partial_u +\frac{c}{12} b\tanh(u)\sinh(u) +\frac{\Delta-\frac{c}{12}}{b}\sech(u)\big]\psi=0,
\end{equation}
with solutions simply modified by an extra phase obtained by integrating the new matter terms,
\begin{equation}
    \psi(b,u) = e^{i\theta(b,u)} \psi_0(b), \qquad \theta(b,u) = \frac{c}{12} b\left(\sinh u-2\tan^{-1}\tanh\tfrac{u}{2}\right)+2\frac{\Delta-\frac{c}{12}}{b}\tan^{-1}\tanh\tfrac{u}{2}.
\end{equation}
In order to determine the group average inner product for these states, we can first determine how time evolution generated by the above Hamiltonian acts on co-invariant geodesic states like\eqref{eq:geostate} by solving the associated time-dependent Schr\"odinger equation. The result is that the inner product is unchanged by coupling to matter, still giving $\li \psi|\psi\ri = \int |\psi_0(b)|^2 b db$. Similar comments apply to other sectors.

\section{Discussion}\label{sec:disc}

\subsection{Hartle-Hawking state}\label{ssec:HH}

Our canonical quantisation gives a new perspective on the Hartle-Hawking no-boundary state in this theory, and in particular the fact that it is non-normalisable (as observed in \cite{Maldacena:2019cbz}).

We can immediately characterise the no-boundary state in our language by thinking about geodesic states \eqref{eq:geostate}, fixing $k=0$ and $a=b$. The corresponding geometry can immediately be continued to a Euclidean section, with metric $ds^2 = d\tau^2 + b^2 \cos^2 \tau d\theta^2$. Locally this is just the metric of a sphere, but at $\tau=-\frac{\pi}{2}$ we will encounter a conical singularity unless $b=1$. The upshot is that the Hartle-Hawking state obtained by a path integral smooth Euclidean metrics without boundary must be proportional to the geodesic state $|b=1\rangle$. In terms of the wavefunctions \eqref{eq:invak} we have
\begin{equation}
    \text{Hartle-Hawking:} \qquad \psi_0(b) \propto  \delta(b-1).
\end{equation}
From this perspective it is not at all surprising that this delta-function wavefunction is not a normalisable state.

We can compare this delta-function divergence to the way in which the sphere partition function diverges \cite{Mahajan:2021nsd}. This can be thought of as arising from dilaton zero modes in the $\ell=1$ mode on the sphere, including $\Phi\propto \sin \tau$ (which corresponds to the $\Phi\propto \sinh t$ classical solutions in Lorentz signature) and two more modes with $\theta$-dependence related by rotations of the sphere. The first of these is captured by the mini-superspace theory, where the magnitude of this solution is canonically conjugate to $b$ (seen in \eqref{eq:phibPoisson}, for example). The wavefunction in terms of this mode is therefore simply the Fourier transform of a delta-function, which does not decay at large $\Phi$ leading to a divergence when we integrate over it. To see the other modes requires us to go beyond the mini-superspace approximation, so a more careful treatment is required, but we see a sign of these modes appearing in the vanishing of the measure in \eqref{eq:muFP}.

\subsection{Quantising  AdS JT closed universes}
We can also apply our analysis to the more familiar AdS JT theory in which we have a negative cosmological constant, albeit still with closed spatial slices.  This is very similar to the dS case, though much less rich. After imposing the gauge conditions $a'=k'=0$ (which in this case captures all states) we are left with a single Hamiltonian constraint
\begin{equation}
    H= -kp-a\Phi.
\end{equation}
If we express this Hamiltonian in the $(a,k)$ representation it becomes 
\begin{equation}
    H= i(k\partial_a -a \partial_k),
\end{equation}
which is simply the generator of rotations in the $(a,k)$ half-plane (as opposed to boosts for the dS version). The physical states (solutions to the WDW equation) will again be states which are constant along these flows. In this case, this means that the physical states are  wavefunctions $\psi(a,k) =\psi_0(\sqrt{a^2 +k^2})$ which are only functions of the `radius' on the $a,k$ plane $b=\sqrt{a^2 +k^2}$ (similar to $b$ and $\beta$ in the de Sitter case). Fixed $b$ corresponds to a fixed metric $ds^2=-dt^2 + b^2 \cos^2 t$, describing a big bang/big crunch cosmology.
%
%
Following the same group averaging procedure as in section \ref{sec:IP&GA} (and applying the same protocol to deal with the non--self-adjoint $H$) we  again get the inner product
\begin{equation}\label{eq:AdSIP}
    \ibraket{\psi}{\psi} = \int_{0}^{\infty} db \,b\,\psi_0^*(b)\psi_0(b)
\end{equation}
on these states. In this case, the wavefunction $\psi_0(b)$ and inner product  \eqref{eq:AdSIP} captures the whole Hilbert space.

\subsection{Many universes and topology changing effects}

Strictly speaking, we have studied not the full phase space and Hilbert space of closed universes, but only the \emph{connected} closed universes. At first sight it is not difficult to remedy this and study all possible universes. The classical phase space (defined by initial data on any compact 1-manifold, for example) simply divides into sectors labelled by a number $N=0,1,2,\ldots$ of universes, where the $N$ sector is the symmetric product of the single-universe phase space we have studied (i.e., the Cartesian product modulo an $S_N$ symmetry that permutes the universes, which can be thought of as a disconnected part of the momentum constraints or spatial  diffeomorphisms). The quantum analogue is that we simply take the Fock space built on the single-universe Hilbert space we have been discussing.

However, this leaves out an interesting possibility: we could have non-perturbative topology-changing effects in the quantum theory, which would couple sectors of different $N$. These have been studied in many papers from the path integral perspective \cite{Maldacena:2019cbz,Cotler_2020,Penington:2019kki}, and we will not add anything new. Instead we make a few speculative comments on how such effects might be incorporated in the Hamiltonian formalism we have used here.

Gravity is special in that non-perturbative effects influence the Hilbert space of the theory, because the Hamiltonian does not generate dynamical evolution but instead acts as a constraint  (except at spatial boundaries). We see this very concretely in the group-averaging proposal for the inner product, which instructs us to integrate over all possible dynamical processes (or a path integral over all possible Lorentzian spacetimes). This leads to a modification of the inner product, which in some regimes may profoundly change the nature of the Hilbert space.

The possibility of such effects has been considered recently in the context of black holes with parametrically large interiors (e.g., black holes formed from collapse after long time evolution), relevant to the information paradox. In particular, the Hilbert space of the interior may have many `null states', meaning that the physical Hilbert space has much smaller dimension than na\"ively expected (from perturbative quantisation around a background classical spacetime, for example) \cite{Penington:2019kki,Marolf:2020xie,Balasubramanian:2022gmo}. The co-invariant states used in section \ref{ssec:coinv} give a language which may be used to explain such null states in terms of gauge invariance. Specifically, we might choose a gauge-fixing condition which specifies a unique wavefunction representing each co-invariant coset in perturbation theory (considering equivalences only under polynomials in the constraints order-by-order in $ \hbar G_N $), but this could leave residual equivalences from `large gauge transformations' which are not close to the identity.

JT gravity is a particularly clean theory to study such effects, because there is a complete separation between `perturbative' physics --- in this case referring to fixed topology, which includes everything in this paper  --- and non-perturbative topology-changing effects suppressed by an independently-tunable parameter $e^{-S_0}$ (in higher dimensions both are controlled by $\hbar G_N$ relative to a typical length scale of the background spacetime).  In our de Sitter theory there is no clear `information paradox', but a potential target to invesigate would be the states of parametrically large universes (with $a\gg 1$), particularly with matter, where one might expect non-perturbative effects to become important and perhaps place some upper bound on the dimension of the physical Hilbert space.

In more realistic theories (with propagating gravitons in particular) one might wonder whether it is fruitful to pursue exponentially small effects when we only have access to perturbation theory around some classical background spacetime. The answer is that the non-perturbative contributions can be qualitatively different to any encountered perturbatively --- in particular, they need not be local in space --- so while small, they are not swamped by perturbative corrections.

Finally, we note that since non-perturbative effects typically require off-shell configurations, they will tend to be excluded in an approach which imposes the constraints before quantisation. They can nonetheless by be captured by the `quantise first' approach that we advocate for. We give simple explicit examples involving quantum tunnelling in \cite{Held:2025mai}. For topology change this is particularly severe, since such processes are incompatible with a smooth Lorentzian geometry. If we wish to retain real Lorentzian metrics, we are required to allow some mild singularities (e.g., of Louko-Sorkin type \cite{Louko:1995jw}). One possible route to this in the Hamiltonian formalism uses the fact that the Hamiltonian constraints are not essentially self-adjoint due to formation of coordinate singularities: perhaps this is not a technical bug, but instead allows us to define the constraints in a way which incorporates such moments of topology change?

\paragraph{Acknowledgements}

We would like to thank Don Marolf, Mukund Rangamani, Douglas Stanford, Lenny Susskind, Sean McBride, David Grabovsky, and Xiaoyi Liu for many helpful discussions. JH is supported by U.S. Department of Defense through the National Defense Science and Engineering Graduate
(NDSEG) Fellowship Program as well as the U.S. Air Force Office of Scientific
Research under award number FA9550-19-1-0360, NSF grant PHY-2408110, and funds from the University of California. HM is supported by DOE grant DE-SC0021085 and a Bloch fellowship from Q-FARM. This research was supported in part by grant NSF PHY-2309135 to the Kavli Institute for Theoretical Physics (KITP).

\appendix

\bibliography{biblio.bib}

\end{document}